\documentclass{easychair}

\usepackage{lineno}
\modulolinenumbers[5]

\usepackage{cite}
\usepackage{textcomp}
\def\BibTeX{{\rm B\kern-.05em{\sc i\kern-.025em b}\kern-.08em
    T\kern-.1667em\lower.7ex\hbox{E}\kern-.125emX}}

\usepackage{xparse}

\usepackage{algorithmic}

\newcommand{\xf}[1]{Figure~\ref{#1}}

\newcommand{\xs}[1]{Section~\ref{#1}}

\newcommand{\xt}[1]{Table~\ref{#1}}

\NewDocumentCommand \styledword{s o d<> m}%
{%
	\IfBooleanTF#1
	{
		\IfNoValueTF{#2}
		{
			\emph{#4}%
		}%
		{
			\emph{\textbf{#4}}%
		}%
	}%
	{
		\IfNoValueTF{#2}
		{
			#4%
		}%
		{
			\textbf{#4}%
		}%
	}%
	\IfValueT{#3}
	{
		\index{#3}%
	}%
}

\NewDocumentCommand \todomsg{o}%
{%
	\begin{center}
		{\color[rgb]{1,0,0} \large \[TODO - {#1}\]}
	\end{center}
}

\NewDocumentCommand \webservice{s o}%
{%
	\def\myword{Web Service}%
	\def\myindex{Web Service}%
	\IfBooleanTF#1
	{
		\IfValueTF{#2}
		{
			\styledword*[#2]<\myindex>{\myword}%
		}%
		{
			\styledword*<\myindex>{\myword}%
		}%
	}%
	{
		\IfValueTF{#2}
		{
			\styledword[#2]<\myindex>{\myword}%
		}%
		{
			\styledword<\myindex>{\myword}%
		}%
	}%
}

\NewDocumentCommand \webservices{s o}%
{%
	\def\myword{Web Services}%
	\def\myindex{Web Service}%
	\IfBooleanTF#1
	{
		\IfValueTF{#2}
		{
			\styledword*[#2]<\myindex>{\myword}%
		}%
		{
			\styledword*<\myindex>{\myword}%
		}%
	}%
	{
		\IfValueTF{#2}
		{
			\styledword[#2]<\myindex>{\myword}%
		}%
		{
			\styledword<\myindex>{\myword}%
		}%
	}%
}

\NewDocumentCommand \service{s o}%
{%
	\def\myword{service}%
	\def\myindex{service}%
	\IfBooleanTF#1
	{
		\IfValueTF{#2}
		{
			\styledword*[#2]<\myindex>{\myword}%
		}%
		{
			\styledword*<\myindex>{\myword}%
		}%
	}%
	{
		\IfValueTF{#2}
		{
			\styledword[#2]<\myindex>{\myword}%
		}%
		{
			\styledword<\myindex>{\myword}%
		}%
	}%
}

\NewDocumentCommand \Service{s o}%
{%
	\def\myword{Service}%
	\def\myindex{service}%
	\IfBooleanTF#1
	{
		\IfValueTF{#2}
		{
			\styledword*[#2]<\myindex>{\myword}%
		}%
		{
			\styledword*<\myindex>{\myword}%
		}%
	}%
	{
		\IfValueTF{#2}
		{
			\styledword[#2]<\myindex>{\myword}%
		}%
		{
			\styledword<\myindex>{\myword}%
		}%
	}%
}

\NewDocumentCommand \services{s o}%
{%
	\def\myword{services}%
	\def\myindex{service}%
	\IfBooleanTF#1
	{
		\IfValueTF{#2}
		{
			\styledword*[#2]<\myindex>{\myword}%
		}%
		{
			\styledword*<\myindex>{\myword}%
		}%
	}%
	{
		\IfValueTF{#2}
		{
			\styledword[#2]<\myindex>{\myword}%
		}%
		{
			\styledword<\myindex>{\myword}%
		}%
	}%
}

\NewDocumentCommand \serviceprovider{s o}%
{%
	\def\myword{service provider}%
	\def\myindex{service provider}%
	\IfBooleanTF#1
	{
		\IfValueTF{#2}
		{
			\styledword*[#2]<\myindex>{\myword}%
		}%
		{
			\styledword*<\myindex>{\myword}%
		}%
	}%
	{
		\IfValueTF{#2}
		{
			\styledword[#2]<\myindex>{\myword}%
		}%
		{
			\styledword<\myindex>{\myword}%
		}%
	}%
}

\NewDocumentCommand \serviceproviders{s o}%
{%
	\def\myword{service providers}%
	\def\myindex{service provider}%
	\IfBooleanTF#1
	{
		\IfValueTF{#2}
		{
			\styledword*[#2]<\myindex>{\myword}%
		}%
		{
			\styledword*<\myindex>{\myword}%
		}%
	}%
	{
		\IfValueTF{#2}
		{
			\styledword[#2]<\myindex>{\myword}%
		}%
		{
			\styledword<\myindex>{\myword}%
		}%
	}%
}

\NewDocumentCommand \Serviceproviders{s o}%
{%
	\def\myword{Service providers}%
	\def\myindex{service provider}%
	\IfBooleanTF#1
	{
		\IfValueTF{#2}
		{
			\styledword*[#2]<\myindex>{\myword}%
		}%
		{
			\styledword*<\myindex>{\myword}%
		}%
	}%
	{
		\IfValueTF{#2}
		{
			\styledword[#2]<\myindex>{\myword}%
		}%
		{
			\styledword<\myindex>{\myword}%
		}%
	}%
}

\NewDocumentCommand \serviceconsumer{s o}%
{%
	\def\myword{service consumer}%
	\def\myindex{service consumer}%
	\IfBooleanTF#1
	{
		\IfValueTF{#2}
		{
			\styledword*[#2]<\myindex>{\myword}%
		}%
		{
			\styledword*<\myindex>{\myword}%
		}%
	}%
	{
		\IfValueTF{#2}
		{
			\styledword[#2]<\myindex>{\myword}%
		}%
		{
			\styledword<\myindex>{\myword}%
		}%
	}%
}

\NewDocumentCommand \serviceconsumers{s o}%
{%
	\def\myword{service consumers}%
	\def\myindex{service consumer}%
	\IfBooleanTF#1
	{
		\IfValueTF{#2}
		{
			\styledword*[#2]<\myindex>{\myword}%
		}%
		{
			\styledword*<\myindex>{\myword}%
		}%
	}%
	{
		\IfValueTF{#2}
		{
			\styledword[#2]<\myindex>{\myword}%
		}%
		{
			\styledword<\myindex>{\myword}%
		}%
	}%
}

\NewDocumentCommand \Serviceconsumers{s o}%
{%
	\def\myword{Service consumers}%
	\def\myindex{service consumer}%
	\IfBooleanTF#1
	{
		\IfValueTF{#2}
		{
			\styledword*[#2]<\myindex>{\myword}%
		}%
		{
			\styledword*<\myindex>{\myword}%
		}%
	}%
	{
		\IfValueTF{#2}
		{
			\styledword[#2]<\myindex>{\myword}%
		}%
		{
			\styledword<\myindex>{\myword}%
		}%
	}%
}

\NewDocumentCommand \endpoint{s o}%
{%
	\def\myword{endpoint}%
	\def\myindex{endpoint}%
	\IfBooleanTF#1
	{
		\IfValueTF{#2}
		{
			\styledword*[#2]<\myindex>{\myword}%
		}%
		{
			\styledword*<\myindex>{\myword}%
		}%
	}%
	{
		\IfValueTF{#2}
		{
			\styledword[#2]<\myindex>{\myword}%
		}%
		{
			\styledword<\myindex>{\myword}%
		}%
	}%
}

\NewDocumentCommand \endpoints{s o}%
{%
	\def\myword{endpoints}%
	\def\myindex{endpoint}%
	\IfBooleanTF#1
	{
		\IfValueTF{#2}
		{
			\styledword*[#2]<\myindex>{\myword}%
		}%
		{
			\styledword*<\myindex>{\myword}%
		}%
	}%
	{
		\IfValueTF{#2}
		{
			\styledword[#2]<\myindex>{\myword}%
		}%
		{
			\styledword<\myindex>{\myword}%
		}%
	}%
}

\NewDocumentCommand \www{s o}%
{%
	\def\myword{World Wide Web}%
	\def\myindex{Web}%
	\IfBooleanTF#1
	{
		\IfValueTF{#2}
		{
			\styledword*[#2]<\myindex>{\myword}%
		}%
		{
			\styledword*<\myindex>{\myword}%
		}%
	}%
	{
		\IfValueTF{#2}
		{
			\styledword[#2]<\myindex>{\myword}%
		}%
		{
			\styledword<\myindex>{\myword}%
		}%
	}%
}

\NewDocumentCommand \web{s o}%
{%
	\def\myword{Web}%
	\def\myindex{Web}%
	\IfBooleanTF#1
	{
		\IfValueTF{#2}
		{
			\styledword*[#2]<\myindex>{\myword}%
		}%
		{
			\styledword*<\myindex>{\myword}%
		}%
	}%
	{
		\IfValueTF{#2}
		{
			\styledword[#2]<\myindex>{\myword}%
		}%
		{
			\styledword<\myindex>{\myword}%
		}%
	}%
}

\NewDocumentCommand \webpage{s o}%
{%
	\def\myword{Web page}%
	\def\myindex{Web page}%
	\IfBooleanTF#1
	{
		\IfValueTF{#2}
		{
			\styledword*[#2]<\myindex>{\myword}%
		}%
		{
			\styledword*<\myindex>{\myword}%
		}%
	}%
	{
		\IfValueTF{#2}
		{
			\styledword[#2]<\myindex>{\myword}%
		}%
		{
			\styledword<\myindex>{\myword}%
		}%
	}%
}

\NewDocumentCommand \webpages{s o}%
{%
	\def\myword{Web pages}%
	\def\myindex{Web page}%
	\IfBooleanTF#1
	{
		\IfValueTF{#2}
		{
			\styledword*[#2]<\myindex>{\myword}%
		}%
		{
			\styledword*<\myindex>{\myword}%
		}%
	}%
	{
		\IfValueTF{#2}
		{
			\styledword[#2]<\myindex>{\myword}%
		}%
		{
			\styledword<\myindex>{\myword}%
		}%
	}%
}

\NewDocumentCommand \internet{s o}%
{%
	\def\myword{Internet}%
	\def\myindex{Internet}%
	\IfBooleanTF#1
	{
		\IfValueTF{#2}
		{
			\styledword*[#2]<\myindex>{\myword}%
		}%
		{
			\styledword*<\myindex>{\myword}%
		}%
	}%
	{
		\IfValueTF{#2}
		{
			\styledword[#2]<\myindex>{\myword}%
		}%
		{
			\styledword<\myindex>{\myword}%
		}%
	}%
}

\NewDocumentCommand \surfaceweb{s o}%
{%
	\def\myword{Surface Web}%
	\def\myindex{Surface Web}%
	\IfBooleanTF#1
	{
		\IfValueTF{#2}
		{
			\styledword*[#2]<\myindex>{\myword}%
		}%
		{
			\styledword*<\myindex>{\myword}%
		}%
	}%
	{
		\IfValueTF{#2}
		{
			\styledword[#2]<\myindex>{\myword}%
		}%
		{
			\styledword<\myindex>{\myword}%
		}%
	}%
}

\NewDocumentCommand \deepweb{s o}%
{%
	\def\myword{Deep Web}%
	\def\myindex{Deep Web}%
	\IfBooleanTF#1
	{
		\IfValueTF{#2}
		{
			\styledword*[#2]<\myindex>{\myword}%
		}%
		{
			\styledword*<\myindex>{\myword}%
		}%
	}%
	{
		\IfValueTF{#2}
		{
			\styledword[#2]<\myindex>{\myword}%
		}%
		{
			\styledword<\myindex>{\myword}%
		}%
	}%
}

\NewDocumentCommand \webcrawler{s o}%
{%
	\def\myword{Web Crawler}%
	\def\myindex{Web Crawler}%
	\IfBooleanTF#1
	{
		\IfValueTF{#2}
		{
			\styledword*[#2]<\myindex>{\myword}%
		}%
		{
			\styledword*<\myindex>{\myword}%
		}%
	}%
	{
		\IfValueTF{#2}
		{
			\styledword[#2]<\myindex>{\myword}%
		}%
		{
			\styledword<\myindex>{\myword}%
		}%
	}%
}

\NewDocumentCommand \webcrawlers{s o}%
{%
	\def\myword{Web Crawlers}%
	\def\myindex{Web Crawler}%
	\IfBooleanTF#1
	{
		\IfValueTF{#2}
		{
			\styledword*[#2]<\myindex>{\myword}%
		}%
		{
			\styledword*<\myindex>{\myword}%
		}%
	}%
	{
		\IfValueTF{#2}
		{
			\styledword[#2]<\myindex>{\myword}%
		}%
		{
			\styledword<\myindex>{\myword}%
		}%
	}%
}

\NewDocumentCommand \seeds{s o}%
{%
	\def\myword{seeds}%
	\def\myindex{Crawling Seeds}%
	\IfBooleanTF#1
	{
		\IfValueTF{#2}
		{
			\styledword*[#2]<\myindex>{\myword}%
		}%
		{
			\styledword*<\myindex>{\myword}%
		}%
	}%
	{
		\IfValueTF{#2}
		{
			\styledword[#2]<\myindex>{\myword}%
		}%
		{
			\styledword<\myindex>{\myword}%
		}%
	}%
}

\NewDocumentCommand \searchengine{s o}%
{%
	\def\myword{search engine}%
	\def\myindex{search engine}%
	\IfBooleanTF#1
	{
		\IfValueTF{#2}
		{
			\styledword*[#2]<\myindex>{\myword}%
		}%
		{
			\styledword*<\myindex>{\myword}%
		}%
	}%
	{
		\IfValueTF{#2}
		{
			\styledword[#2]<\myindex>{\myword}%
		}%
		{
			\styledword<\myindex>{\myword}%
		}%
	}%
}

\NewDocumentCommand \searchengines{s o}%
{%
	\def\myword{search engines}%
	\def\myindex{search engine}%
	\IfBooleanTF#1
	{
		\IfValueTF{#2}
		{
			\styledword*[#2]<\myindex>{\myword}%
		}%
		{
			\styledword*<\myindex>{\myword}%
		}%
	}%
	{
		\IfValueTF{#2}
		{
			\styledword[#2]<\myindex>{\myword}%
		}%
		{
			\styledword<\myindex>{\myword}%
		}%
	}%
}

\NewDocumentCommand \Searchengines{s o}%
{%
	\def\myword{Search engines}%
	\def\myindex{search engine}%
	\IfBooleanTF#1
	{
		\IfValueTF{#2}
		{
			\styledword*[#2]<\myindex>{\myword}%
		}%
		{
			\styledword*<\myindex>{\myword}%
		}%
	}%
	{
		\IfValueTF{#2}
		{
			\styledword[#2]<\myindex>{\myword}%
		}%
		{
			\styledword<\myindex>{\myword}%
		}%
	}%
}

\NewDocumentCommand \repository{s o}%
{%
	\def\myword{repository}%
	\def\myindex{repository}%
	\IfBooleanTF#1
	{
		\IfValueTF{#2}
		{
			\styledword*[#2]<\myindex>{\myword}%
		}%
		{
			\styledword*<\myindex>{\myword}%
		}%
	}%
	{
		\IfValueTF{#2}
		{
			\styledword[#2]<\myindex>{\myword}%
		}%
		{
			\styledword<\myindex>{\myword}%
		}%
	}%
}

\NewDocumentCommand \Repository{s o}%
{%
	\def\myword{Repository}%
	\def\myindex{repository}%
	\IfBooleanTF#1
	{
		\IfValueTF{#2}
		{
			\styledword*[#2]<\myindex>{\myword}%
		}%
		{
			\styledword*<\myindex>{\myword}%
		}%
	}%
	{
		\IfValueTF{#2}
		{
			\styledword[#2]<\myindex>{\myword}%
		}%
		{
			\styledword<\myindex>{\myword}%
		}%
	}%
}

\NewDocumentCommand \repositories{s o}%
{%
	\def\myword{repositories}%
	\def\myindex{repository}%
	\IfBooleanTF#1
	{
		\IfValueTF{#2}
		{
			\styledword*[#2]<\myindex>{\myword}%
		}%
		{
			\styledword*<\myindex>{\myword}%
		}%
	}%
	{
		\IfValueTF{#2}
		{
			\styledword[#2]<\myindex>{\myword}%
		}%
		{
			\styledword<\myindex>{\myword}%
		}%
	}%
}

\NewDocumentCommand \Repositories{s o}%
{%
	\def\myword{Repositories}%
	\def\myindex{repository}%
	\IfBooleanTF#1
	{
		\IfValueTF{#2}
		{
			\styledword*[#2]<\myindex>{\myword}%
		}%
		{
			\styledword*<\myindex>{\myword}%
		}%
	}%
	{
		\IfValueTF{#2}
		{
			\styledword[#2]<\myindex>{\myword}%
		}%
		{
			\styledword<\myindex>{\myword}%
		}%
	}%
}

\NewDocumentCommand \registry{s o}%
{%
	\def\myword{registry}%
	\def\myindex{registry}%
	\IfBooleanTF#1
	{
		\IfValueTF{#2}
		{
			\styledword*[#2]<\myindex>{\myword}%
		}%
		{
			\styledword*<\myindex>{\myword}%
		}%
	}%
	{
		\IfValueTF{#2}
		{
			\styledword[#2]<\myindex>{\myword}%
		}%
		{
			\styledword<\myindex>{\myword}%
		}%
	}%
}

\NewDocumentCommand \uddi{s o}%
{%
	\def\myword{UDDI}%
	\def\myindex{UDDI}%
	\IfBooleanTF#1
	{
		\IfValueTF{#2}
		{
			\styledword*[#2]<\myindex>{\myword}%
		}%
		{
			\styledword*<\myindex>{\myword}%
		}%
	}%
	{
		\IfValueTF{#2}
		{
			\styledword[#2]<\myindex>{\myword}%
		}%
		{
			\styledword<\myindex>{\myword}%
		}%
	}%
}

\NewDocumentCommand \uddidef{s o}%
{%
	\def\myword{Universal Description, Discovery and Integration}%
	\def\myindex{UDDI}%
	\IfBooleanTF#1
	{
		\IfValueTF{#2}
		{
			\styledword*[#2]<\myindex>{\myword}%
		}%
		{
			\styledword*<\myindex>{\myword}%
		}%
	}%
	{
		\IfValueTF{#2}
		{
			\styledword[#2]<\myindex>{\myword}%
		}%
		{
			\styledword<\myindex>{\myword}%
		}%
	}%
}

\NewDocumentCommand \servicediscovery{s o}%
{%
	\def\myword{Service Discovery}%
	\def\myindex{Service Discovery}%
	\IfBooleanTF#1
	{
		\IfValueTF{#2}
		{
			\styledword*[#2]<\myindex>{\myword}%
		}%
		{
			\styledword*<\myindex>{\myword}%
		}%
	}%
	{
		\IfValueTF{#2}
		{
			\styledword[#2]<\myindex>{\myword}%
		}%
		{
			\styledword<\myindex>{\myword}%
		}%
	}%
}

\NewDocumentCommand \snapshot{s o}%
{%
	\def\myword{snapshot}%
	\def\myindex{snapshot}%
	\IfBooleanTF#1
	{
		\IfValueTF{#2}
		{
			\styledword*[#2]<\myindex>{\myword}%
		}%
		{
			\styledword*<\myindex>{\myword}%
		}%
	}%
	{
		\IfValueTF{#2}
		{
			\styledword[#2]<\myindex>{\myword}%
		}%
		{
			\styledword<\myindex>{\myword}%
		}%
	}%
}

\NewDocumentCommand \Snapshot{s o}%
{%
	\def\myword{Snapshot}%
	\def\myindex{snapshot}%
	\IfBooleanTF#1
	{
		\IfValueTF{#2}
		{
			\styledword*[#2]<\myindex>{\myword}%
		}%
		{
			\styledword*<\myindex>{\myword}%
		}%
	}%
	{
		\IfValueTF{#2}
		{
			\styledword[#2]<\myindex>{\myword}%
		}%
		{
			\styledword<\myindex>{\myword}%
		}%
	}%
}

\NewDocumentCommand \snapshots{s o}%
{%
	\def\myword{snapshots}%
	\def\myindex{snapshot}%
	\IfBooleanTF#1
	{
		\IfValueTF{#2}
		{
			\styledword*[#2]<\myindex>{\myword}%
		}%
		{
			\styledword*<\myindex>{\myword}%
		}%
	}%
	{
		\IfValueTF{#2}
		{
			\styledword[#2]<\myindex>{\myword}%
		}%
		{
			\styledword<\myindex>{\myword}%
		}%
	}%
}

\NewDocumentCommand \Snapshots{s o}%
{%
	\def\myword{Snapshots}%
	\def\myindex{snapshot}%
	\IfBooleanTF#1
	{
		\IfValueTF{#2}
		{
			\styledword*[#2]<\myindex>{\myword}%
		}%
		{
			\styledword*<\myindex>{\myword}%
		}%
	}%
	{
		\IfValueTF{#2}
		{
			\styledword[#2]<\myindex>{\myword}%
		}%
		{
			\styledword<\myindex>{\myword}%
		}%
	}%
}
\NewDocumentCommand \soa{s o}%
{%
	\def\myword{SOA}%
	\def\myindex{SOA}%
	\IfBooleanTF#1
	{
		\IfValueTF{#2}
		{
			\styledword*[#2]<\myindex>{\myword}%
		}%
		{
			\styledword*<\myindex>{\myword}%
		}%
	}%
	{
		\IfValueTF{#2}
		{
			\styledword[#2]<\myindex>{\myword}%
		}%
		{
			\styledword<\myindex>{\myword}%
		}%
	}%
}

\NewDocumentCommand \soadef{s o}%
{%
	\def\myword{Service-Oriented Architecture}%
	\def\myindex{SOA}%
	\IfBooleanTF#1
	{
		\IfValueTF{#2}
		{
			\styledword*[#2]<\myindex>{\myword}%
		}%
		{
			\styledword*<\myindex>{\myword}%
		}%
	}%
	{
		\IfValueTF{#2}
		{
			\styledword[#2]<\myindex>{\myword}%
		}%
		{
			\styledword<\myindex>{\myword}%
		}%
	}%
}

\NewDocumentCommand \iaas{s o}%
{%
	\def\myword{IaaS}%
	\def\myindex{IaaS}%
	\IfBooleanTF#1
	{
		\IfValueTF{#2}
		{
			\styledword*[#2]<\myindex>{\myword}%
		}%
		{
			\styledword*<\myindex>{\myword}%
		}%
	}%
	{
		\IfValueTF{#2}
		{
			\styledword[#2]<\myindex>{\myword}%
		}%
		{
			\styledword<\myindex>{\myword}%
		}%
	}%
}

\NewDocumentCommand \iaasdef{s o}%
{%
	\def\myword{Infrastructure as a Service}%
	\def\myindex{IaaS}%
	\IfBooleanTF#1
	{
		\IfValueTF{#2}
		{
			\styledword*[#2]<\myindex>{\myword}%
		}%
		{
			\styledword*<\myindex>{\myword}%
		}%
	}%
	{
		\IfValueTF{#2}
		{
			\styledword[#2]<\myindex>{\myword}%
		}%
		{
			\styledword<\myindex>{\myword}%
		}%
	}%
}

\NewDocumentCommand \paas{s o}%
{%
	\def\myword{PaaS}%
	\def\myindex{PaaS}%
	\IfBooleanTF#1
	{
		\IfValueTF{#2}
		{
			\styledword*[#2]<\myindex>{\myword}%
		}%
		{
			\styledword*<\myindex>{\myword}%
		}%
	}%
	{
		\IfValueTF{#2}
		{
			\styledword[#2]<\myindex>{\myword}%
		}%
		{
			\styledword<\myindex>{\myword}%
		}%
	}%
}

\NewDocumentCommand \paasdef{s o}%
{%
	\def\myword{Platform as a Service}%
	\def\myindex{PaaS}%
	\IfBooleanTF#1
	{
		\IfValueTF{#2}
		{
			\styledword*[#2]<\myindex>{\myword}%
		}%
		{
			\styledword*<\myindex>{\myword}%
		}%
	}%
	{
		\IfValueTF{#2}
		{
			\styledword[#2]<\myindex>{\myword}%
		}%
		{
			\styledword<\myindex>{\myword}%
		}%
	}%
}

\NewDocumentCommand \saas{s o}%
{%
	\def\myword{SaaS}%
	\def\myindex{SaaS}%
	\IfBooleanTF#1
	{
		\IfValueTF{#2}
		{
			\styledword*[#2]<\myindex>{\myword}%
		}%
		{
			\styledword*<\myindex>{\myword}%
		}%
	}%
	{
		\IfValueTF{#2}
		{
			\styledword[#2]<\myindex>{\myword}%
		}%
		{
			\styledword<\myindex>{\myword}%
		}%
	}%
}

\NewDocumentCommand \saasdef{s o}%
{%
	\def\myword{Software as a Service}%
	\def\myindex{SaaS}%
	\IfBooleanTF#1
	{
		\IfValueTF{#2}
		{
			\styledword*[#2]<\myindex>{\myword}%
		}%
		{
			\styledword*<\myindex>{\myword}%
		}%
	}%
	{
		\IfValueTF{#2}
		{
			\styledword[#2]<\myindex>{\myword}%
		}%
		{
			\styledword<\myindex>{\myword}%
		}%
	}%
}

\NewDocumentCommand \cloud{s o}%
{%
	\def\myword{Cloud}%
	\def\myindex{Cloud}%
	\IfBooleanTF#1
	{
		\IfValueTF{#2}
		{
			\styledword*[#2]<\myindex>{\myword}%
		}%
		{
			\styledword*<\myindex>{\myword}%
		}%
	}%
	{
		\IfValueTF{#2}
		{
			\styledword[#2]<\myindex>{\myword}%
		}%
		{
			\styledword<\myindex>{\myword}%
		}%
	}%
}

\NewDocumentCommand \clouddef{s o}%
{%
	\def\myword{Cloud Computing}%
	\def\myindex{Cloud}%
	\IfBooleanTF#1
	{
		\IfValueTF{#2}
		{
			\styledword*[#2]<\myindex>{\myword}%
		}%
		{
			\styledword*<\myindex>{\myword}%
		}%
	}%
	{
		\IfValueTF{#2}
		{
			\styledword[#2]<\myindex>{\myword}%
		}%
		{
			\styledword<\myindex>{\myword}%
		}%
	}%
}

\NewDocumentCommand \broker{s o}%
{%
	\def\myword{Broker}%
	\def\myindex{Broker}%
	\IfBooleanTF#1
	{
		\IfValueTF{#2}
		{
			\styledword*[#2]<\myindex>{\myword}%
		}%
		{
			\styledword*<\myindex>{\myword}%
		}%
	}%
	{
		\IfValueTF{#2}
		{
			\styledword[#2]<\myindex>{\myword}%
		}%
		{
			\styledword<\myindex>{\myword}%
		}%
	}%
}

\NewDocumentCommand \brokerage{s o}%
{%
	\def\myword{Brokerage}%
	\def\myindex{Brokerage}%
	\IfBooleanTF#1
	{
		\IfValueTF{#2}
		{
			\styledword*[#2]<\myindex>{\myword}%
		}%
		{
			\styledword*<\myindex>{\myword}%
		}%
	}%
	{
		\IfValueTF{#2}
		{
			\styledword[#2]<\myindex>{\myword}%
		}%
		{
			\styledword<\myindex>{\myword}%
		}%
	}%
}

\NewDocumentCommand \sla{s o}%
{%
	\def\myword{SLA}%
	\def\myindex{SLA}%
	\IfBooleanTF#1
	{
		\IfValueTF{#2}
		{
			\styledword*[#2]<\myindex>{\myword}%
		}%
		{
			\styledword*<\myindex>{\myword}%
		}%
	}%
	{
		\IfValueTF{#2}
		{
			\styledword[#2]<\myindex>{\myword}%
		}%
		{
			\styledword<\myindex>{\myword}%
		}%
	}%
}

\NewDocumentCommand \sladef{s o}%
{%
	\def\myword{Service Level Agreements}%
	\def\myindex{SLA}%
	\IfBooleanTF#1
	{
		\IfValueTF{#2}
		{
			\styledword*[#2]<\myindex>{\myword}%
		}%
		{
			\styledword*<\myindex>{\myword}%
		}%
	}%
	{
		\IfValueTF{#2}
		{
			\styledword[#2]<\myindex>{\myword}%
		}%
		{
			\styledword<\myindex>{\myword}%
		}%
	}%
}

\NewDocumentCommand \context{s o}%
{%
	\def\myword{context}%
	\def\myindex{Context}%
	\IfBooleanTF#1
	{
		\IfValueTF{#2}
		{
			\styledword*[#2]<\myindex>{\myword}%
		}%
		{
			\styledword*<\myindex>{\myword}%
		}%
	}%
	{
		\IfValueTF{#2}
		{
			\styledword[#2]<\myindex>{\myword}%
		}%
		{
			\styledword<\myindex>{\myword}%
		}%
	}%
}

\NewDocumentCommand \Context{s o}%
{%
	\def\myword{Context}%
	\def\myindex{Context}%
	\IfBooleanTF#1
	{
		\IfValueTF{#2}
		{
			\styledword*[#2]<\myindex>{\myword}%
		}%
		{
			\styledword*<\myindex>{\myword}%
		}%
	}%
	{
		\IfValueTF{#2}
		{
			\styledword[#2]<\myindex>{\myword}%
		}%
		{
			\styledword<\myindex>{\myword}%
		}%
	}%
}

\NewDocumentCommand \contextual{s o}%
{%
	\def\myword{contextual}%
	\def\myindex{Context}%
	\IfBooleanTF#1
	{
		\IfValueTF{#2}
		{
			\styledword*[#2]<\myindex>{\myword}%
		}%
		{
			\styledword*<\myindex>{\myword}%
		}%
	}%
	{
		\IfValueTF{#2}
		{
			\styledword[#2]<\myindex>{\myword}%
		}%
		{
			\styledword<\myindex>{\myword}%
		}%
	}%
}

\NewDocumentCommand \Contextual{s o}%
{%
	\def\myword{Contextual}%
	\def\myindex{Context}%
	\IfBooleanTF#1
	{
		\IfValueTF{#2}
		{
			\styledword*[#2]<\myindex>{\myword}%
		}%
		{
			\styledword*<\myindex>{\myword}%
		}%
	}%
	{
		\IfValueTF{#2}
		{
			\styledword[#2]<\myindex>{\myword}%
		}%
		{
			\styledword<\myindex>{\myword}%
		}%
	}%
}

\NewDocumentCommand \contextaware{s o}%
{%
	\def\myword{context-aware}%
	\def\myindex{Context}%
	\IfBooleanTF#1
	{
		\IfValueTF{#2}
		{
			\styledword*[#2]<\myindex>{\myword}%
		}%
		{
			\styledword*<\myindex>{\myword}%
		}%
	}%
	{
		\IfValueTF{#2}
		{
			\styledword[#2]<\myindex>{\myword}%
		}%
		{
			\styledword<\myindex>{\myword}%
		}%
	}%
}

\NewDocumentCommand \serviceclass{s o}%
{%
	\def\myword{Service Classification}%
	\def\myindex{Service Classification}%
	\IfBooleanTF#1
	{
		\IfValueTF{#2}
		{
			\styledword*[#2]<\myindex>{\myword}%
		}%
		{
			\styledword*<\myindex>{\myword}%
		}%
	}%
	{
		\IfValueTF{#2}
		{
			\styledword[#2]<\myindex>{\myword}%
		}%
		{
			\styledword<\myindex>{\myword}%
		}%
	}%
}

\NewDocumentCommand \machinelearning{s o}%
{%
	\def\myword{Machine Learning}%
	\def\myindex{Machine Learning}%
	\IfBooleanTF#1
	{
		\IfValueTF{#2}
		{
			\styledword*[#2]<\myindex>{\myword}%
		}%
		{
			\styledword*<\myindex>{\myword}%
		}%
	}%
	{
		\IfValueTF{#2}
		{
			\styledword[#2]<\myindex>{\myword}%
		}%
		{
			\styledword<\myindex>{\myword}%
		}%
	}%
}

\NewDocumentCommand \marfdef{s o}%
{%
	\def\myword{Modular Audio Recognition Framework}%
	\def\myindex{MARF}%
	\IfBooleanTF#1
	{
		\IfValueTF{#2}
		{
			\styledword*[#2]<\myindex>{\myword}%
		}%
		{
			\styledword*<\myindex>{\myword}%
		}%
	}%
	{
		\IfValueTF{#2}
		{
			\styledword[#2]<\myindex>{\myword}%
		}%
		{
			\styledword<\myindex>{\myword}%
		}%
	}%
}

\NewDocumentCommand \marf{s o}%
{%
	\def\myword{MARF}%
	\def\myindex{MARF}%
	\IfBooleanTF#1
	{
		\IfValueTF{#2}
		{
			\styledword*[#2]<\myindex>{\myword}%
		}%
		{
			\styledword*<\myindex>{\myword}%
		}%
	}%
	{
		\IfValueTF{#2}
		{
			\styledword[#2]<\myindex>{\myword}%
		}%
		{
			\styledword<\myindex>{\myword}%
		}%
	}%
}

\NewDocumentCommand \marfcat{s o}%
{%
	\def\myword{MARFCAT}%
	\def\myindex{MARFCAT}%
	\IfBooleanTF#1
	{
		\IfValueTF{#2}
		{
			\styledword*[#2]<\myindex>{\myword}%
		}%
		{
			\styledword*<\myindex>{\myword}%
		}%
	}%
	{
		\IfValueTF{#2}
		{
			\styledword[#2]<\myindex>{\myword}%
		}%
		{
			\styledword<\myindex>{\myword}%
		}%
	}%
}

\NewDocumentCommand \marfcatdef{s o}%
{%
	\def\myword{MARF-based Code Analysis Tool}%
	\def\myindex{MARFCAT}%
	\IfBooleanTF#1
	{
		\IfValueTF{#2}
		{
			\styledword*[#2]<\myindex>{\myword}%
		}%
		{
			\styledword*<\myindex>{\myword}%
		}%
	}%
	{
		\IfValueTF{#2}
		{
			\styledword[#2]<\myindex>{\myword}%
		}%
		{
			\styledword<\myindex>{\myword}%
		}%
	}%
}

\NewDocumentCommand \mca{s o}%
{%
	\def\myword{MCA}%
	\def\myindex{MCA}%
	\IfBooleanTF#1
	{
		\IfValueTF{#2}
		{
			\styledword*[#2]<\myindex>{\myword}%
		}%
		{
			\styledword*<\myindex>{\myword}%
		}%
	}%
	{
		\IfValueTF{#2}
		{
			\styledword[#2]<\myindex>{\myword}%
		}%
		{
			\styledword<\myindex>{\myword}%
		}%
	}%
}

\NewDocumentCommand \mcadef{s o}%
{%
	\def\myword{Mean Class Accuracy}%
	\def\myindex{MCA}%
	\IfBooleanTF#1
	{
		\IfValueTF{#2}
		{
			\styledword*[#2]<\myindex>{\myword}%
		}%
		{
			\styledword*<\myindex>{\myword}%
		}%
	}%
	{
		\IfValueTF{#2}
		{
			\styledword[#2]<\myindex>{\myword}%
		}%
		{
			\styledword<\myindex>{\myword}%
		}%
	}%
}

\NewDocumentCommand \xml{s o}%
{%
	\def\myword{XML}%
	\def\myindex{XML}%
	\IfBooleanTF#1
	{
		\IfValueTF{#2}
		{
			\styledword*[#2]<\myindex>{\myword}%
		}%
		{
			\styledword*<\myindex>{\myword}%
		}%
	}%
	{
		\IfValueTF{#2}
		{
			\styledword[#2]<\myindex>{\myword}%
		}%
		{
			\styledword<\myindex>{\myword}%
		}%
	}%
}

\NewDocumentCommand \soap{s o}%
{%
	\def\myword{SOAP}%
	\def\myindex{SOAP}%
	\IfBooleanTF#1
	{
		\IfValueTF{#2}
		{
			\styledword*[#2]<\myindex>{\myword}%
		}%
		{
			\styledword*<\myindex>{\myword}%
		}%
	}%
	{
		\IfValueTF{#2}
		{
			\styledword[#2]<\myindex>{\myword}%
		}%
		{
			\styledword<\myindex>{\myword}%
		}%
	}%
}

\NewDocumentCommand \soapdef{s o}%
{%
	\def\myword{Simple Object Access Protocol}%
	\def\myindex{SOAP}%
	\IfBooleanTF#1
	{
		\IfValueTF{#2}
		{
			\styledword*[#2]<\myindex>{\myword}%
		}%
		{
			\styledword*<\myindex>{\myword}%
		}%
	}%
	{
		\IfValueTF{#2}
		{
			\styledword[#2]<\myindex>{\myword}%
		}%
		{
			\styledword<\myindex>{\myword}%
		}%
	}%
}

\NewDocumentCommand \wsdl{s o}%
{%
	\def\myword{WSDL}%
	\def\myindex{WSDL}%
	\IfBooleanTF#1
	{
		\IfValueTF{#2}
		{
			\styledword*[#2]<\myindex>{\myword}%
		}%
		{
			\styledword*<\myindex>{\myword}%
		}%
	}%
	{
		\IfValueTF{#2}
		{
			\styledword[#2]<\myindex>{\myword}%
		}%
		{
			\styledword<\myindex>{\myword}%
		}%
	}%
}

\NewDocumentCommand \wsdldef{s o}%
{%
	\def\myword{Web Service Description Language}%
	\def\myindex{WSDL}%
	\IfBooleanTF#1
	{
		\IfValueTF{#2}
		{
			\styledword*[#2]<\myindex>{\myword}%
		}%
		{
			\styledword*<\myindex>{\myword}%
		}%
	}%
	{
		\IfValueTF{#2}
		{
			\styledword[#2]<\myindex>{\myword}%
		}%
		{
			\styledword<\myindex>{\myword}%
		}%
	}%
}

\NewDocumentCommand \restful{s o}%
{%
	\def\myword{RESTful}%
	\def\myindex{REST}%
	\IfBooleanTF#1
	{
		\IfValueTF{#2}
		{
			\styledword*[#2]<\myindex>{\myword}%
		}%
		{
			\styledword*<\myindex>{\myword}%
		}%
	}%
	{
		\IfValueTF{#2}
		{
			\styledword[#2]<\myindex>{\myword}%
		}%
		{
			\styledword<\myindex>{\myword}%
		}%
	}%
}

\NewDocumentCommand \restdef{s o}%
{%
	\def\myword{Representational State Transfer}%
	\def\myindex{REST}%
	\IfBooleanTF#1
	{
		\IfValueTF{#2}
		{
			\styledword*[#2]<\myindex>{\myword}%
		}%
		{
			\styledword*<\myindex>{\myword}%
		}%
	}%
	{
		\IfValueTF{#2}
		{
			\styledword[#2]<\myindex>{\myword}%
		}%
		{
			\styledword<\myindex>{\myword}%
		}%
	}%
}

\NewDocumentCommand \rest{s o}%
{%
	\def\myword{REST}%
	\def\myindex{REST}%
	\IfBooleanTF#1
	{
		\IfValueTF{#2}
		{
			\styledword*[#2]<\myindex>{\myword}%
		}%
		{
			\styledword*<\myindex>{\myword}%
		}%
	}%
	{
		\IfValueTF{#2}
		{
			\styledword[#2]<\myindex>{\myword}%
		}%
		{
			\styledword<\myindex>{\myword}%
		}%
	}%
}

\NewDocumentCommand \wadl{s o}%
{%
	\def\myword{WADL}%
	\def\myindex{WADL}%
	\IfBooleanTF#1
	{
		\IfValueTF{#2}
		{
			\styledword*[#2]<\myindex>{\myword}%
		}%
		{
			\styledword*<\myindex>{\myword}%
		}%
	}%
	{
		\IfValueTF{#2}
		{
			\styledword[#2]<\myindex>{\myword}%
		}%
		{
			\styledword<\myindex>{\myword}%
		}%
	}%
}

\NewDocumentCommand \wadldef{s o}%
{%
	\def\myword{Web Application Description Language}%
	\def\myindex{WADL}%
	\IfBooleanTF#1
	{
		\IfValueTF{#2}
		{
			\styledword*[#2]<\myindex>{\myword}%
		}%
		{
			\styledword*<\myindex>{\myword}%
		}%
	}%
	{
		\IfValueTF{#2}
		{
			\styledword[#2]<\myindex>{\myword}%
		}%
		{
			\styledword<\myindex>{\myword}%
		}%
	}%
}

\NewDocumentCommand \http{s o}%
{%
	\def\myword{HTTP}%
	\def\myindex{HTTP}%
	\IfBooleanTF#1
	{
		\IfValueTF{#2}
		{
			\styledword*[#2]<\myindex>{\myword}%
		}%
		{
			\styledword*<\myindex>{\myword}%
		}%
	}%
	{
		\IfValueTF{#2}
		{
			\styledword[#2]<\myindex>{\myword}%
		}%
		{
			\styledword<\myindex>{\myword}%
		}%
	}%
}

\NewDocumentCommand \httpdef{s o}%
{%
	\def\myword{Hyper Text Transfer Protocol}%
	\def\myindex{HTTP}%
	\IfBooleanTF#1
	{
		\IfValueTF{#2}
		{
			\styledword*[#2]<\myindex>{\myword}%
		}%
		{
			\styledword*<\myindex>{\myword}%
		}%
	}%
	{
		\IfValueTF{#2}
		{
			\styledword[#2]<\myindex>{\myword}%
		}%
		{
			\styledword<\myindex>{\myword}%
		}%
	}%
}

\NewDocumentCommand \urlword{s o}%
{%
	\def\myword{URL}%
	\def\myindex{URL}%
	\IfBooleanTF#1
	{
		\IfValueTF{#2}
		{
			\styledword*[#2]<\myindex>{\myword}%
		}%
		{
			\styledword*<\myindex>{\myword}%
		}%
	}%
	{
		\IfValueTF{#2}
		{
			\styledword[#2]<\myindex>{\myword}%
		}%
		{
			\styledword<\myindex>{\myword}%
		}%
	}%
}

\NewDocumentCommand \urls{s o}%
{%
	\def\myword{URLs}%
	\def\myindex{URL}%
	\IfBooleanTF#1
	{
		\IfValueTF{#2}
		{
			\styledword*[#2]<\myindex>{\myword}%
		}%
		{
			\styledword*<\myindex>{\myword}%
		}%
	}%
	{
		\IfValueTF{#2}
		{
			\styledword[#2]<\myindex>{\myword}%
		}%
		{
			\styledword<\myindex>{\myword}%
		}%
	}%
}

\NewDocumentCommand \html{s o}%
{%
	\def\myword{HTML}%
	\def\myindex{HTML}%
	\IfBooleanTF#1
	{
		\IfValueTF{#2}
		{
			\styledword*[#2]<\myindex>{\myword}%
		}%
		{
			\styledword*<\myindex>{\myword}%
		}%
	}%
	{
		\IfValueTF{#2}
		{
			\styledword[#2]<\myindex>{\myword}%
		}%
		{
			\styledword<\myindex>{\myword}%
		}%
	}%
}

\NewDocumentCommand \java{s o}%
{%
	\def\myword{Java}%
	\def\myindex{Java}%
	\IfBooleanTF#1
	{
		\IfValueTF{#2}
		{
			\styledword*[#2]<\myindex>{\myword}%
		}%
		{
			\styledword*<\myindex>{\myword}%
		}%
	}%
	{
		\IfValueTF{#2}
		{
			\styledword[#2]<\myindex>{\myword}%
		}%
		{
			\styledword<\myindex>{\myword}%
		}%
	}%
}

\newcommand{\option}[1]{\texttt{#1}\index{Options!#1}}

\newcommand{\lucidL}[1]{{$\mathit{Lucid}$}($L$) }

		{}

\def\myvert{\raise 2.27pt \hbox{\vrule depth 0pt height 8pt width 0.2mm}}
\def\myarrow{\hspace*{0.43mm}%
             \raise 2.29pt\hbox{\vrule depth 0pt height 8pt width 0.16mm}%
             \hspace*{-0.32mm}%
             $\longrightarrow$
             \ %
             }

\begin{document}


\title{Fast Context-Annotated Classification of Different Types of Web Service Descriptions}
\titlerunning{MARFCAT-based Web Service Classifications}



\author
{
Serguei A. Mokhov, Joey Paquet, Arash Khodadadi\\
\affiliation{Department of Computer Science and Software Engineering} \\
\affiliation{Concordia University}\\
\affiliation{Montreal, Canada}\\
\affiliation{\url{{mokhov,paquet,ar_khoda}@cse.concordia.ca}}
}
\authorrunning{Mokhov, Paquet, Khodadadi}

\maketitle

\begin{abstract}
In the recent rapid growth of web services, IoT, and cloud computing,
many web services and APIs appeared on the web. With the failure of
global UDDI registries, different service repositories started to
appear, trying to list and categorize various types of web services
for client applications' discover and use.

In order to increase the effectiveness and speed up the task of finding %
compatible \webservices{} in the brokerage when performing \service{} %
composition or suggesting \webservices{} to the requests, high-level %
functionality of the \service{} needs to be determined.
Due to the lack of structured support for specifying such functionality, %
classification of \services{} into a set of abstract categories is necessary.

We employ a wide range of \machinelearning{} and Signal Processing algorithms %
and techniques in order to find the highest precision achievable in the scope %
of this article for the fast classification of three type of \service{} descriptions:
\wsdl{}, \rest{}, and \wadl{}.
In addition, we complement our approach by showing the importance and effect %
of \contextual{} information on the classification of the \service{} %
descriptions and show that it improves the accuracy in 5 different categories of \services{}.
%
\end{abstract}

%

\section{Introduction}
\label{sec:Introduction}

When two systems (e.g. \webservices) want to interact, a compatibility assessment %
which requires in-depth analysis considering the interface and %
conversational protocol of them needs to take place.
As the authors in %
\cite{machine-learning-ws-description-classification-2012} argue, one way %
to speed up the assessment above is to apply \machinelearning{} methods to %
automatically classify high-level functionality of a system's interface %
description, i.e, the highest level of abstraction of what the system does.
This will result in restricting the scope of compatibility checks and %
consequently providing an overall performance gain when looking for matches %
between systems.
In addition to increasing performance of compatibility assessment, 
the authors in %
\cite{clustering-facilitate-service-discovery-springer-2014} argue that %
classifying \webservices{} into different sets based on the tags (clustering) %
facilitates the task of \servicediscovery{}.
Moreover, the result of \serviceclass{} can be very useful to the end-users %
when selecting \services{}.

\serviceclass{} or Categorization is the task of associating \webservice{} %
descriptions to a predefined set of categories which can considerably speed %
up and increase the effectiveness of the task of finding compatible %
\webservices{} in \brokerage{} or suggesting \webservices{} to the requests %
\cite{machine-learning-ws-description-classification-2012}.
%
Categories or classes specify the purpose of the \service{} and what it does %
at a high level.
However, there is no structured support for specifying the %
abstract category to which the service belongs \cite{machine-learning-ws-description-classification-2012}.
As a result, this classification task needs to be done manually or %
automatically.

There are two main approaches towards \serviceclass{}: %
\emph{manual classification} and \emph{automatic classification}.
According to \cite{crowd-sourcing-service-annotations-AAAI-2012}, the former %
methods are very expensive, both in time, effort and consequently financially.
The latter methods however are quite inaccurate and do not in general provide %
quality annotations but cheaper than the former.
Although the authors in %
\cite{crowd-sourcing-service-annotations-AAAI-2012} try to %
decrease the cost of manual classification by applying crowd sourcing %
techniques, it is still more expensive than the automatic methods.  
%
Due to the fact that cost plays an important role and because of the %
resources available to us, although we are aware that the annotations in many %
cases are inaccurate and the automatic classification may not be as accurate %
as manual methods, we use automatic classification approach as our primary %
methodology.
We build on the considerable amount of research that has been carried %
out on the topic of automatic classification of a text document which has %
many practical applications~\cite{automatic-text-categorization-2005}.

The task of automatic classification of documents is usually tackled by %
applying \machinelearning*{} techniques. 
These techniques use classifiers that have been automatically %
induced by estimation on a collection of documents which is called the %
\emph{training set} \cite{mitchell2002}.
\machinelearning{} methods can be divided into two broad categories: (1) {\it supervised learning}, where each document in the training set is already associated with a category by %
a human supervisor and (2) {\it unsupervised learning}, where documents are not associated with a category prior to the learning process %
and the \machinelearning{} method must find a meaningful division into %
categories. In this article we focus on the former method, which has generally been much %
more successful in most studies as pointed out in
\cite{machine-learning-ws-description-classification-2012}.

\Context{}, in the \webservices{} environment is any %
information about the \serviceconsumer{}, \serviceprovider{}, and %
communication protocols.
Hence, content of the \service{} descriptions and any information related to %
them is considered as a \context{} for the \service{}.

In \cite{khodadadi-mcthesis2015}
we discussed harvesting and storing \webservice{} %
descriptions and their \contextual{} information from different sources.
Eventually, we found 72,454 unique \service{} description \urlword s %
including 39,288 \wsdl{} \urlword s, 1,830 \wadl{} \urlword s, and 31,336 %
\html{} page \urlword s describing \restful{} \services{}.
From these \urlword s we stored 48,161 actual \service{} description files %
including 16,096 \wsdl{} descriptions, 450 \wadl{} descriptions, and 31,615 %
\html{} files describing \restful{} \services{}.
We constructed a repository of \webservice{} descriptions and their \contextual{} information.
In this article we try to find the highest accuracy achievable in the scope of %
this article by employing a wide range of \machinelearning*{} and \emph{Signal %
Processing} algorithms and techniques and putting the \context*{} %
into practice.

We use the open-source \marf{} framework and its \marfcat{} application %
because they are designed as an input media type-independent investigation platform %
to execute a considerable number of experiments in a short amount of time and %
to assist selecting the best combinations of different available algorithms.
In this application, we use signal processing techniques which use character-%
level (bi-grams) processing rather than syntax and semantic levels and we treat the %
descriptions as a \emph{signal} which will be discussed in details in %
\xs{sec:MARF}.
In \xs{sec:MARFAlgorithms} and \xs{sec:MARFOptions} we discuss different %
algorithms and options available in \marfcat{}.

Using \marfcat{} as our investigation platform, we systematically test and %
select the best (a tradeoff between accuracy, recall, and speed) %
combination(s) of algorithm implementations (configuration) available to %
us for each type of \service{} descriptions (\xs{sec:DataState}) and then use %
only those for the final classification of all \service{} descriptions based %
on the classes defined in \xs{sec:Classes}.
We will discuss our methodology in \xs{sec:TestingMethodology}.

\section{Background}

\subsection{MARF and MARFCAT}
\label{sec:MARF}


\marfdef*{} (\marf*{})~\cite{marf}, is an open-source %
collection of pattern recognition APIs and their implementation for %
unsupervised and supervised \machinelearning{} and classification.
\marf{} was designed to act as a testbed to verify and test common and novel %
algorithms found in literature for sample loading, pre-processing, feature %
extraction, and training and classification, which constitute a %
typical pattern recognition pipeline \cite{marf-c3s2e08}.
Over the years, \marf{} accumulated a fair %
number of implementations for each of the pipeline stages which allows us to %
execute reasonably comprehensive comparative studies of algorithm %
combinations for the \serviceclass{} purpose.
%
%

The pattern recognition process starts by \textbf{loading} a sample (e.g., an %
audio recording, a text, or image file), %
removing noisy and/or silent data and other unwanted elements %
(\textbf{pre-processing}), %
then extracting the most prominent features from it
(\textbf{feature extraction}), %
and finally either \textbf{training} %
the system such that the system learns a new set of features of a given %
subject or \textbf{classifying} what the subject is.
The outcome of the training process is either a collection of some form of %
feature vectors or their mean or median clusters, which are stored for every %
subject learned.
The outcome of classification is the class that the system believes the %
subject belongs to and a score attached to it~\cite{marf-c3s2e08}.



The loading stage in \marf{} starts with the interpretation of the files being %
scanned in terms of bytes forming amplitude values in a signal %
using either unigram, bi-gram, or tri-gram approach. 
Then, the pre-processing allows to be \emph{none-at-all} (raw, or the fastest), %
\emph{normalization}, traditional frequency domain filters, wavelet-based filters, %
etc. 
Feature extraction involves reducing an arbitrary length signal to a %
fixed length feature vector of what is thought to be the most relevant %
features in the signal, e.g., spectral features in FFT, LPC, min-max %
amplitudes, etc.
The classification stage is then separated either to train by learning the %
incoming feature vectors (usually $k$-means clusters, median clusters, or plain %
feature vector collection, combined with neural network training) or testing %
them against the previously learned models \cite{marfcat-sate4-arxiv}.

%

\marfcat*{} is a \marfcatdef{}, which was first %
exhibited at the Static Analysis Tool Exposition (SATE) workshop in 2010 %
\cite{nist-samate-sate2010,nist-samate-sate4}.
\marfcat{}, as any {\marf} application, can be used for a wide array of recognition tasks, not only applicable %
to audio, but rather to general pattern recognition for various applications, %
such as in digital forensic analysis, writer identification, natural language %
processing (NLP) \cite{marf-nlp-framework}, among others.
In particular, \marfcat{} was used to analyze source and byte code to fingerprint, detect, and
classify vulnerabilities and weaknesses in %
\cite{marfcat-sate2010-nist,marfcat-signal-swan2015,marfcat-nlp-ai2014} and do the same for network
packet traces \cite{mal-traffic-classification-dpi-headers}.
The authors point out \marfcat{}'s advantages and shortcomings:

\subsection*{Advantages}
	\begin{itemize}
		\item Relatively fast (e.g., 2400 files to train and test in about 3 minutes) %
 on a now-commodity 7-year old desktop or a laptop.
		\item Input data type-independent %
		(e.g., sound files, binary and source code, images, and natural language text)
		\item Can automatically learn a large knowledge-base to test on known and %
		unknown cases.
		\item A wide range of algorithms and their combinations can be %
investigated to select the best ones for a particular task.
	\end{itemize}
\subsection*{Shortcomings}
	\begin{itemize}
		\item Interpreting a signal is less intuitive by humans in the output.
		\item Accuracy depends on the quality of the knowledge-base %
		(training sets) collected.
		Some of this collection and annotation is manual; hence, error-prone and a %
subject to over-fitting.
	\end{itemize}



\subsection{Motivation to Use MARFCAT}

The following are primary motivations justifying the use of {\marfcat}
in this work:

\begin{enumerate}
	\item 
{\marfcat} was successfully used in related source code and text analysis tasks, for specific 
vulnerabilities and defects as well as more general weakness categories as referenced earlier.
At its introduction in 2010, it was arguably the first time such an approach
was applied to text analysis and was deemed novel in these types of tasks.
The most significant advantage of it was the processing speed compared to other
code analysis tools. By extension this applied it to \webservices{} descriptions in various formats.

	\item 
{\marfcat} supports both signal processing and NLP pipelines.
However, the signal pipeline was found by an order of magnitude
faster than most parsing and NLP approaches~\cite{marfcat-nlp-ai2014,marfcat-signal-swan2015}. 
Thus, spectral analysis was proven beneficial in code analysis, source, and binary 
as well as network packet traces, and natural language processing.
It is analogous to analyze the signal from a distant star, breaking it 
down into spectrum of emitted light in order to classify the chemical 
composition in terms of elements present in the star, i.e., to fingerprint
them.
{\marfcat} similarly fingerprints a spectrum of text or any other media into 
bins related to different categories it was shown to learn from.

	\item
{\marfcat} is very easy to quickly setup and do preliminary testing
in search for good algorithms. It can also be used as a front-end
for semantic- and ontology-based parsing classifiers to prioritize
their work~\cite{marfcat-sate2010-nist}.

	\item 
{\marfcat}'s author was readily available to consult on issues of its uses
and operation.

\end{enumerate}


\section{Methodology}
\label{sect:methodology}

\subsection{Architecture}
\label{sec:Architecture}

The architecture of this entire work is illustrated in \xf{fig:architecture}.
This article focuses predominantly on service classification block
illustrated in this figure.
Specifically, \xf{fig:architecture} illustrates the whole architecture and how the two %
steps of \service{} collecting and \serviceclass{} are connected.


\begin{figure}[htbp]
    \centering
        \includegraphics[width=.7\columnwidth]{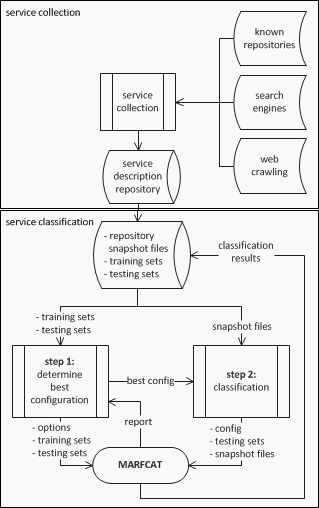}
    \caption{Service collection/classification architecture}
    \label{fig:architecture}
\end{figure}

\xf{fig:architecture} depicts the \serviceclass{} %
concrete architecture and components.
As discussed in \xs{sec:TestingMethodology}, initially we test each sample %
type (\xs{sec:DataState}) independently in order to find the best %
configuration of \marfcat{} for that sample type.

At the end, after finding the best configuration for each sample type, we %
create the testing sets for each type from all the \service{} descriptions in %
the \repository{} which are not classified yet.
Then we perform the final classification for each type based on the best %
configurations (algorithm combination + clustering method) found from the %
previous step and store the resulting classes associated with each %
\snapshot{} in the \snapshots{} info part of the \service{} descriptions %
\repository{}.


\subsection{Data Sets, Classes, and Training Sets}
\label{sec:DataModel}
\label{sec:DataState}
\label{sec:Classes}

Our \service{} \repository{} consists of two main \repositories{}: (1) the {\bf \Service{} 
Descriptions \Repository{}} stores \service{} description \urlword s, providers, \contextual{} %
information, and \snapshots{} information and references. This is the 
main repository structure and serves 
as a basis for the generation of our data sets (snapshots) for the 
purpose of the current study. This \repository{} is implemented as a SQL database. 
(2) The {\bf \Snapshots{} \Repository{}}, is a file-based \repository{} which stores \snapshots{} %
of \service{} descriptions, \context{} files, and \snapshots{} of \service{} %
descriptions combined with their \context{} files. Each \snapshot{} is stored in a folder named with its \serviceprovider{} %
\urlword{} and linked to \Snapshots{} information in the \service{} descriptions %
\repository{}.

%



In \cite{khodadadi-mcthesis2015}, we have designed a web crawler to search the web for 
web service descriptions in order to create an web service repository able to store 
as many web service descriptions that we could find. In doing so, we have identified the 
three main service descriptions commonly used: \wsdl{}, \wadl{} and \rest. 
In order to convey the current study, we store each service description and their \context{} separately.
Due to the nature of each type's description's characteristics and features, %
and to compare and analyze the results separately, %
we survey each service description type's classification process independently.

For \wsdl{} and \wadl{} files, we use the descriptions directly, which we found %
using our web crawler.
On the other hand, for the service descriptions regarding \rest{} %
\services{}, which are often described using \html{} files, we take an additional step before feeding them to \marfcat{} %
because of the nature of these files which contain too much noise, e.g., %
using \emph{scripts} to strip the code.

In this step we remove all the \html{} tags and unnecessary sections and only %
keep the raw text inside and store it in a separate text file and consider it %
as a new type of sample.
Likewise, this step can be applied to \wsdl{} and \wadl{} files to remove all %
their tags.
However, they do not contain much noise and \marfcat{} will take care %
of noise removal in the pre-processing step.
Therefore, this task were postponed to the future work because it will %
multiply the number of the tests to be applied.

As a result, our \repository{} will contain four general types of samples:

\begin{itemize}
	\item \wsdl{} files (.wsdl)
	\item \wadl{} files (.wadl)
	\item \html{} files (.html)
	\item Tags-Filtered description files (.txt)
\end{itemize}

Another dimension which is added to each of these types is their %
\contextual{} information.
In order to show the effect of \context{} on the classification and to find %
the best configuration, we define three type of samples with respect to the %
\contextual{} information for each of general types defined above:

\begin{itemize}
	\item Plain files (description files without any \context{} added to them)
	\item Combined with \context{} files (plain descriptions + \context{})
	\item Only \context{} files (files containing only the \contextual{} %
information of \service{} descriptions)
\end{itemize}
 

Each of these data sets are then to be loaded into \marf{} and processed following the data flow 
shown in \xf{fig:classification-dataflow}.



We use 5 classes for the classification with respect to previous %
research 
\cite{titan-web-service-discovery-www12-2012}, %
\cite{tag-relevance-service-mining-springer-2014}, %
\cite{distributed-QoS-evaluation-for-real-world-services-IEEE-2010}, %
the most popular categories in ProgrammableWeb\footnote{\url{http://%
www.programmableweb.com/}} and more importantly the nature of \webservice{} %
descriptions in the \repository{} and their intersections:

\begin{itemize}
	\item Weather
	\item Social
	\item Tourism
	\item Entertainment
	\item Financial
\end{itemize}

As mentioned in \xs{sec:MARF}, in order to classify the \service{} %
descriptions which are stored in the \repository{}, we need \emph{training %
sets} for each of these classes.
These sets need to be chosen with minimal intersections to be definite %
candidates for the class.
In addition, for the testing purposes to find the best combination of %
configurations which will be discussed in \xs{sec:TestingMethodology}, we %
need \emph{testing sets} for each of the classes.
Therefore, we manually classify 500 instances (100 per each class) for %
\wsdl{} and \rest{} files.
However, for \wadl{} files because of the inadequacy of the files in the %
\repository{} as discussed in \cite[Chapter 3]{khodadadi-mcthesis2015}, %
only for weather, social and financial 10 definite matches can %
be found.
For tourism 3 candidates and for entertainment only 2 candidates are chosen.

\subsection{MARFCAT Configuration}
\label{sec:MARFconfig}



We use \marf{} and its %
application \marfcat{} to find the best algorithm combinations for each %
sample type which were mentioned in \xs{sec:DataState}.
We use the \emph{fast} script of \marfcat{} which performs the algorithms %
illustrated in \xf{fig:classification-dataflow} in each step of its pipeline: %
1 Loader, 4 techniques in the Preparation stage, 9 algorithms in the Pre-%
processing stage, 4 algorithms in the Feature Extraction stage, and 6 %
Distance Classifier algorithms. The combination of these algorithms will result in 864 permutations which we %
test in each of our cases as discussed in \xs{sec:TestingMethodology}. The following is a brief description of some 
of the algorithms and options that we have used in this research. 

\begin{figure}[htbp]
	\centering
		\includegraphics[width=\columnwidth
		]%
		{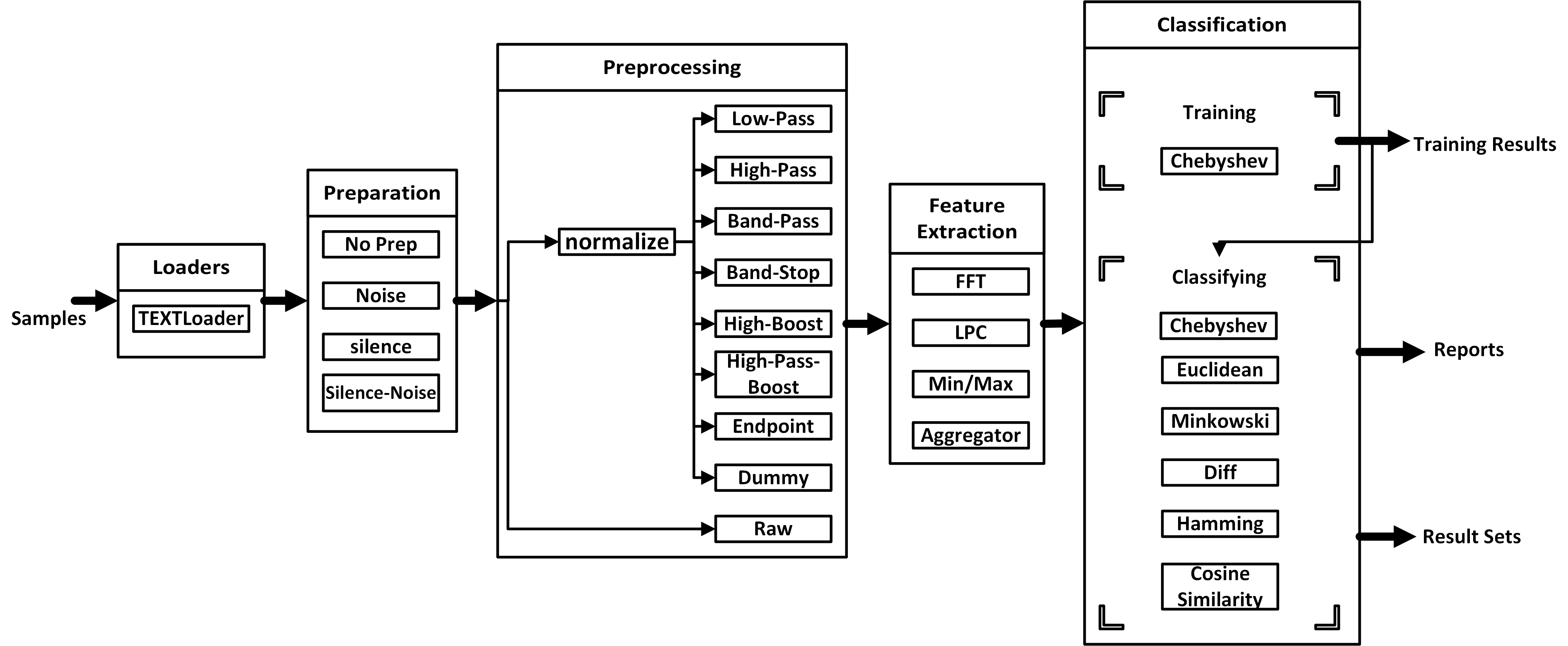}
	\caption{Service classification data flow in MARF}
	\label{fig:classification-dataflow}
\end{figure}

\subsection*{Algorithms Used by MARFCAT}
\label{sec:MARFAlgorithms}

The specific algorithms come from the classical literature and other sources %
and are detailed in \cite{marf-c3s2e08}, %
\cite{marf}, %
\cite{mal-traffic-classification-dpi-headers}.
The below is a summary of some algorithms corresponding to %
\cite{marf-c3s2e08} with a brief description:

\noindent{\bf Fast Fourier Transform (FFT): }
	A version of the Discrete Fourier Transform used %
in FFT-based filtering as well as feature extraction \cite{dspdimension}.
It is also used in FFT-based filters (both forward and inverse FFT
to reconstruct the signal after filtering).
Uses 512 frequencies by default (empirically determined by the MARF project).

\noindent{\bf Linear Predictive Coding (LPC): }
  Used in feature extraction, %
which evaluates windowed sections of signals and determines a %
set of coefficients approximating the amplitude vs.\ frequency function.
Uses 20 poles by default.

\noindent{\bf Distance Classifiers: }
Various distance classifiers (Chebyshev, Euclidean, and Minkowski \cite{abdi-%
distance}, Mahalanobis \cite{mahalanobis-distance}, Diff (internally %
developed within \marf{}, roughly similar in behavior to the UNIX/Linux diff %
utility \cite{diff}), and Hamming \cite{hamming-distance}).

\noindent{\bf Cosine Similarity Measure: }
Cosine similarity measure \cite{cosine-similarity-tutorial}, \cite{cosine-%
similarity-euclidean-distance} was thoroughly discussed in %
\cite{khalifemcthesis04} and often produces the best or near best accuracy in %
\marf{} in many configurations.

\subsection*{MARFCAT Options}
\label{sec:MARFOptions}

All mentioned algorithms are selected as options in a scripted
manner exhaustively at the first stage in order to select the
candidate best options for subsequent classification.
Not all combinations are necessarily optimal or have effect
together (e.g., noise removal uses low pass filter at pre-processing and then
if low pass filter is applied, it doubles the work, without additional filtering effect),
but they are easy to automate and there is no dependency assumptions between
algorithms at different stages keeping them decoupled and re-usable.
%
We survey some of these options to find the best configuration for each type of %
\service{} description which will be discussed in \xs{sec:TestingMethodology}.

In order to be able to classify samples into different classes, an automatic %
classifier determines the salient properties of the samples and puts them %
into different feature vectors. 
This process is called feature extraction \cite{machine-learning-ws-%
description-classification-2012}.
In \marf{} there are different ways of storing and matching feature %
vectors that {\marfcat} takes advantages of from a specific class.
These are referred to as clustering options in %
\marfcat{} and can be customized:

\begin{itemize}
	\item k-means clusters (mean option)
	\item median clusters (median option)
	\item plain feature vector collection (no clustering option)
\end{itemize}

We report our results with all three clustering options to find the best configuration for %
each type of \service{} description which will be discussed in %
\xs{sec:TestingMethodology}.


In terms of some most prominent algorithms producing the best results
in the algorithm selection stage include, but not limited to:


\subsection*{Preparation stage options: }

\noindent{\bf \option{-noise}} does noise removal
by applying an FFT (Fast Fourier Transform) low-pass filter effectively removing
high-frequency occurring material.

\noindent{\bf \option{-silence}} removes near-zero gaps from the data.
It is important to apply silence removal after the noise removal since
noise filtering may produce more silence gaps. The gaps are removed by
compression of the input data into a smaller sized array
by cutting out and concatenating non-silent portions.

\noindent{\bf \option{-silence-noise}} combines the noise and silence removal. It helped
selecting best low-frequency non-zero local minimums and maximum features
in classification of less structured samples such as \rest{} descriptions.
\subsection*{Preprocessing stage options}

\noindent{\bf \option{-endp}} is endpointing which collects all \emph{local} minimums and maximums from the signal.
It worked best with \option{-minmax}.

\noindent{\bf \option{-low}} FFT filter that removes approximately the upper $1/3$ band
of frequency spectrum by applying a zero frequency response on that portion
effectively removing most high-frequency bigram material. It is redundant to apply \option{-low} and \option{-noise} together
under the current implementation of \option{-noise}, but scripting facilities
do not make such intelligent guesses.

\noindent{\bf \option{-bandstop}} keeps approximately the lowest and highest $1/3$ bands of the spectrum and removing
the middle third. Combined together with the low-pass filter this effectively
means $2/3$ upper frequencies are removed keeping $1/3$ of the lower-frequency
band.
%
\subsection*{Feature Extraction stage options}

\noindent{\bf \option{-minmax}} picks a hundred features from the vector,
where 50 are minimums and 50 are maximums. If there are less than 100 values,
the gap is filled with zeros. Worked best with \option{-endp} to select 
the 100 local minimum and maximums.

\noindent{\bf \option{-lpc}} is Linear Predictive Coding which works on a spectral envelope of
coefficients representing the spectrum curve. In \marf{}, the empirical default
is 20 coefficients. It works well with compressed form of signal, such as 
with local minimums and maximums with silence removed. The feature vectors
are as a result small -- 20 features making distance calculation faster
(as opposed to \option{-fft}'s 512 features).

\subsection*{Classification stage options}
		\begin{itemize}
		\item 

\noindent{\bf \option{-cheb}} is Chebyshev distance classifier which appears to work best
with the \option{-endp} and \option{-minmax} selected local extremes
due to their block nature
that provides enough discriminatory power for highly varied and overlapping
data sets such as \rest{} descriptions.
It is also the fastest classifier.
		\item

\noindent{\bf \option{-eucl}} is Euclidean distance which works better with less
varied 20-sized vectors, such as produced by LPC combined with endpointing.
\option{-silence}, \option{-endp}, \option{-lpc} with Euclidean distance
appear to produce one of the best configurations during search for algorithms
to use for \wsdl{} descriptions.
		\end{itemize}

\begin{itemize}
	\item 
Preprocessing: \option{-endp} -- endpointing collects all \emph{local} minimums and maximums from the signal.
It worked best with \option{-minmax}.

	\item 
Feature extraction: \option{-minmax} -- picks hundred features from the vector,
where 50 are minimums and 50 are maximums. If there are less than 100 values,
the gap is filled with zeros. Worked best with \option{-endp} to select 
the 100 local minimum and maximums.

	\item 
Feature extraction: \option{-lpc} -- LPC works on a spectral envelope of
coefficients representing the spectrum curve. In MARF, the empirical default
is 20 of them. It works well with compressed form of signal, such as 
with local minimums and maximums with silence removed. The feature vectors
are as a result small -- 20 features making distance calculation faster
(as opposed to \option{-fft}'s 512).

	\item 
Pre-Preprocessing: \option{-noise} currently instructs to do noise removal
by applying as an FFT low-pass filter effectively removing
high-frequency occurring material.

	\item 
Pre-Preprocessing: \option{-silence} removes near-zero gaps from the data.
It is important to apply silence removal after the noise removal since
noise filtering may produce more silence gaps. The gaps are removed by
compression of the input data into a smaller sized array
by cutting out and concatenating non-silent portions.

	\item 
Prepocessing: \option{-low} pass FFT filter removes approximately upper $1/3$ band
of frequency spectrum by applying a zero frequency response on that portion
effectively removing most high-frequency bigram material.

(It is redundant to apply \option{-low} and \option{-noise} together
under the current implementation of \option{-noise}, but scripting facilities
do not make such intelligent guesses.)

	\item 
Preprocessing: \option{-bandstop} keeps approximately the lowest and highest $1/3$ bands of the spectrum and removing
the middle third. Combined together with the low-pass filter this effectively
means $2/3$ upper frequencies are removed keeping $1/3$ of the lower-frequency
band.

This appears to have contribute well to the to algorithms where mid-range bigrams
were also a part of the textual noise. As a corollary, that means perhaps
unigrams would not require a bandstop and perhaps a single lowpass would
be sufficient.


\item
Classification: \option{-cheb} -- Chebyshev distance classifier appears to work best
with the \option{-endp} and \option{-minmax} selected local extremes
due to their block nature
that provides enough discriminatory power for highly varied and overlapping
RESTful services.
It is also the fastest classifier.

\item
Classification: \option{-eucl} -- Euclidean distance works better with less
varied 20-sized vectors, such as produced by LPC combined with endpointing.
\option{-silence}, \option{-endp}, \option{-lpc} with Euclidean distance
appear to produce one of the best configurations during search for algorithms
to use for WSDLs.

\end{itemize}

Other options and their algorithms, and their complexity are discussed
in \cite{marf-c3s2e08,mal-traffic-classification-dpi-headers} and related
and are omitted here for brevity.


\subsection{Testing Methodology}
\label{sec:TestingMethodology}

As discussed before in \xs{sec:MARF}, we use \marf{} and its application %
\marfcat{} to find the best algorithm combinations for each \service{} %
description types which were mentioned in \xs{sec:DataState}.
In order to perform this task, \marfcat{} defines two processes which were %
discussed in \xs{sec:MARF}: (1) The {\it learning process} in which the feature vectors are extracted and the system learns the %
classes from the \emph{training set} and (2) the {\it testing} or {\it classification process} in which the \emph{testing set} is classified based on the %
previously learned models.

In order to evaluate the performance of the classifiers, %
we compute different evaluation measures.
These measures are usually presented as percentages. 
Consider for a given class $C$, $n_{t}$ samples are expected, i.e., are labeled with $C$.
The classification system classifies (labels) $n_{s}$ samples as $C$ including $n_{c}$ correct samples (\emph{true positives}) and $n_{n}$ incorrect samples (\emph{false positives}).

\noindent{\bf Total Accuracy: } Total accuracy is defined as the fraction of the samples %
which were classified in the same class as expected in total:
$$\frac{\sum n_{c}}{\sum n_{t}}$$

\noindent{\bf Precision: } Precision is defined for each class as the fraction of classified items which %
are relevant, i.e., expected in that class:
$$\frac{n_{c}}{n_{s}}$$
We also compute the \emph{macro precision} which is the average of precision %
over all classes.

\noindent{\bf Recall: } Recall (also called sensitivity) is defined for each class as the fraction %
of relevant items which are classified:
$$\frac{n_{c}}{n_{t}}$$
We also compute the \emph{macro recall} which is the average of recall %
over all classes.

\noindent{\bf F-Measure: } F-Measure is defined for each class as the harmonic mean of precision and %
recall:
$$2 * \frac{precision * recall}{precision + recall}$$
We also compute the \emph{macro F-Measure} which is the average of f-measure %
over all classes.

\noindent{\bf Classification Time: } Classification time is the total execution time of the classification process over the data set.

In our methodology, initially we survey each description type %
(\xs{sec:DataState}) independently in order to find the best algorithm %
combinations considering the above-mentioned measurements. In addition, there are also other options which %
\marfcat{} provides and were referred in \xs{sec:MARFOptions} as clustering %
and frequency options.
We test all three clustering options in all cases to find the best %
algorithm combinations.
On the other hand, because the frequency change did not have any effect on %
the precision (see \xs{sec:ClassificationResults}) when tested in the best case, we %
ignored it for the other cases. Therefore, we find the best configuration of \marfcat{} for each sample %
types which consists of an algorithm combination using a specific clustering %
method.

As argued in \xs{sec:Classes}, we chose 5 classes and we manually %
classified 500 instances (100 per class) for each \service{} description type.
As discussed in \xs{sec:DataState}, \contextual{} information adds %
another dimension to the samples and adds 2 more sample types (plain + %
\context{}, and only \context{}) for each of the \service{} description types. In order to completely survey the possible cases and find the best %
configuration(s), we train and test on all type of samples exhaustively.

\xf{fig:testing-methodology} illustrates our methodology which %
forms 72 different cases based on the sample types and clustering options.
Each case consists of one row from each of the blocks; one description type, training and testing on which type of file considering \contextual{} information. 
We tested 864 algorithm permutations in the Signal Processing pipeline for each %
case
and %
illustrated in \xf{fig:classification-dataflow}.

\begin{figure*}[htbp]
	\centering
		\includegraphics[width=\textwidth]
		{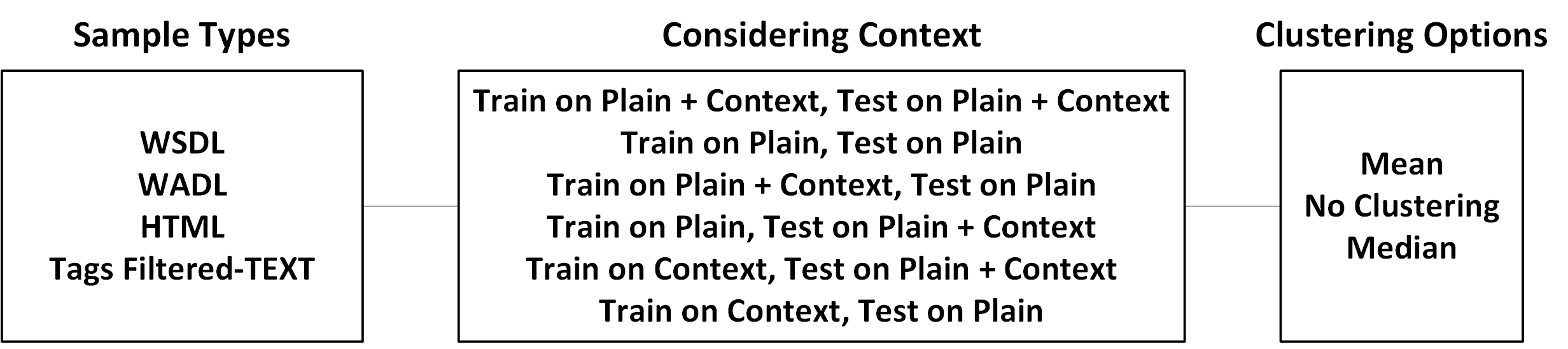}
	\caption{Testing Methodology}
	\label{fig:testing-methodology}
\end{figure*}

In this exhaustive test process we choose randomly a smaller set of the %
manually classified instances in order to keep the tests simple and %
applicable in a shorter amount of time (62,208 runs). After finding the best cases, we increase the sets and use all 500 instances %
for each of \service{} description types.
Using these sets we perform another exhaustive search on the algorithm %
combinations in order to find the best algorithm combination.

Using the best case and the best algorithm combination, we perform a 10-fold %
cross-validation in order to give an insight on how the model will generalize %
to an independent dataset and to reduce variability.
In this procedure, we split the data randomly into 10 pieces and run the %
classification 10 times using one of the pieces (10\%) as the testing set and %
the rest (90\%) as the training set in a way that each sample is present once %
and only once in the testing set among the runs and then, average all the %
results.

In order to show the effect of the use of \context{} on the classification, %
we study the best configuration without any \contextual{} information added, %
and the best one through all other cases with \context{}.
As a result, we perform the 10-fold cross-validation for each of these cases %
and compare the results in order to illustrate the effect of \context{}.
We then compare the evaluation measures (total accuracy, macro recall and %
precision, macro F-Measure, and the classification time) in order to show the %
effect of \context{}.
In addition, we compare the performance of our classification for \wsdl{} %
files with the literature in order to give an insight on how close our %
classification stands.
However, we could not find any related work for classification of WADL or \rest{} %
descriptions to compare.

At the end, after finding the best configuration for each sample type, we %
create the sets from all the \service{} descriptions in the %
\repository{} which are not classified yet and perform the classification for %
them and store the results in the \repository{}.

\section{Results and Evaluation}
\label{sec:ClassificationResults}
\index{Results and Evaluation}

As discussed in \xs{sec:TestingMethodology}, in order to classify the %
\service{} descriptions, we first find the best configuration of \marfcat{} %
(best algorithm combination + clustering option).
In addition, we add the \contextual{} information to the classification, our hypothesis being that
%
adding \contextual{} information to the files will improve the performance %
of the classification.

Using this approach, as illustrated in \xs{sec:TestingMethodology}, based on the %
sample types and considering the \contextual{} information and the clustering %
options, we survey 72 different cases.
As illustrated in %
\xf{fig:classification-dataflow}, we exhaustively test 864 algorithm %
permutations for each case.
As discussed in \xs{sec:DataState}, we classify and survey each \service{} %
description type individually.
For each \service{} type, we %
survey 18 different cases in order to find the highest accuracy achievable %
without considering any \contextual{} information (training on \emph{Plain} %
files and testing on \emph{Plain} files) and with the \contextual{} %
information in effect (e.g., training on \emph{Plain + \Context{}} files and %
testing on \emph{Plain + \Context{}} files).


\subsection{WSDL Classification Results}
\label{sec:WSDLsClassificationResults}

\xt{tab:WSDLsCasesPrecision} depicts the highest accuracy achievable %
by exhaustively testing 864 algorithm permutations for each case of \wsdl{} %
files using different clustering options which were mentioned in %
\xs{sec:MARFOptions}.

\begin{table}[htbp]
\centering
\caption{Classification accuracy across data sets for \wsdl{} files}
\label{tab:WSDLsCasesPrecision}
\resizebox{\textwidth}{!}{%
\begin{tabular}{|c|c|c|c|c|c|c|}
\hline
 & {\bf \begin{tabular}[c]{@{}c@{}}TrainPlainCtx\\ TestPlainCtx\end{tabular}} & {\bf \begin{tabular}[c]{@{}c@{}}TrainPlain\\ TestPlain\end{tabular}} & {\bf \begin{tabular}[c]{@{}c@{}}TrainPlainCtx\\ TestPlain\end{tabular}} & {\bf \begin{tabular}[c]{@{}c@{}}TrainPlain\\ TestPlainCtx\end{tabular}} & {\bf \begin{tabular}[c]{@{}c@{}}TrainCtx\\ TestPlainCtx\end{tabular}} & {\bf \begin{tabular}[c]{@{}c@{}}TrainCtx\\ TestPlain\end{tabular}} \\ \hline
{\bf Mean} & 60  & 52 & 48 & 52 & 36 & 28 \\ \hline
{\bf No Clustering} & 68 & {\bf 64} &{\bf 76} & 64 & 36 & 36 \\ \hline
{\bf Median} & 56 & 56 & 52 & 52 & 28 & 32 \\ \hline
\end{tabular}
}
\end{table}

The highest accuracy without considering \context{} (Train on Plain-Test on %
Plain column) is achieved by using the \textit{No Clustering} option: %
64 percent.
The highest accuracy with considering \context{} (other columns) is achieved %
by using the \emph{No Clustering} option and using \emph{Plain + \Context{}} %
files in the training and \emph{Plain} files in the testing: 76 percent.


As discussed in \xs{sec:TestingMethodology}, after finding the best case %
which is using the \emph{No Clustering} option and using \emph{Plain + %
\Context{}} files in the training and \emph{Plain} files in the testing, we %
increase the sets and perform another exhaustive search in order to find the %
best algorithm combination. 

As the results depicted, the best result for the classification of \wsdl{} %
files is achieved by training on \emph{Plain + \Context{}} files and testing %
on \emph{Plain} files using the following configuration:

\begin{itemize}
	\item No Clustering option (discussed in \xs{sec:MARFOptions})
	\item \option{-silence} for preparation, \option{-endp} for pre-processing, \option{-lpc} for feature extraction, \option{-eucl} for classification
\end{itemize}

As mentioned in \xs{sec:MARFOptions}, \option{-silence} option removes near-zero gaps from the data.
There are usually many white-spaces and empty parts in the \wsdl{} files that are normalized close to
zero or silence gaps appear due to low-pass filtering.
As a result this preparation technique helped to improve the overall classification combination.
Additionally, as mentioned in \xs{sec:MARFOptions} from a theoretical point of view, 
LPC works well with compressed form of signal, such as with local minimums 
and maximums with silence removed.
%
In addition, Euclidean distance (which is sensitive to high-dimensional vectors)
works better with less varied 20-sized vectors, such as produced by LPC combined with endpointing.

As discussed in \xs{sec:TestingMethodology}, we perform 10-fold cross-%
validation based on this configuration in order to give an insight on how the %
model will generalize to an independent dataset and to reduce variability.
\xt{tab:WSDLsClassificationComparison} depicts the cross-validated results including %
the evaluation measures which were defined in \xs{sec:TestingMethodology}.


We compare the performance of our classification for \wsdl{} files with the %
literature in order to give an insight on how close our classification %
process is to the literature.
However, most of the research has been carried out for semantically-defined %
files (using Ontology Language (OWL-S) \cite{owl-s} and Web Service Modeling %
Ontology (WSMO) \cite{wsmo}) which are not available at a large scale and are %
a small subset of available \service{} description files.
The authors in \cite{machine-learning-ws-description-classification-2012} %
used different techniques for feature extraction such as Bag of Words %
variances and different algorithms for machine learning such as Support %
Vector Machines (SVM) variances and compared them in order to find the best %
classification performance for \wsdl{} files.
Finally, the best combination was the result of employing Support Vector %
Machines and use a feature extractor that is tailored to the task of \wsdl{} %
classification by using its structure, in particular the identifiers.
\xt{tab:WSDLsClassificationComparison} (first row) depicts the results of their work including %
the same evaluation measures except the classification time, which is not %
presented as part of their results.


Although the data sets from which the tests are performed are different, we can %
conclude that our classification accuracy is very close to the best %
result from the literature without any customization on the preprocessing, feature %
extraction, and classification based on the \wsdl{} files.
%
{\marfcat} offers a good tradeoff
between precision and speed and helps us to validate the hypothesis
on the positive effect of context on classification results.

In order to show the effect of \contextual{} information, as discussed in %
\xs{sec:TestingMethodology}, we perform another 10-fold cross-validation on %
the same configuration without considering any \contextual{} information and %
training on the \emph{Plain} files and testing on the \emph{Plain} files.
\xt{tab:WSDLsClassificationComparison} (second row) depicts the cross-validated results %
including the evaluation measures without considering any \contextual{} %
information.

\begin{table}[htbp]
\centering
\caption{Effect of context on cross-validated \wsdl{} classification results}
\label{tab:WSDLsClassificationComparison}
\resizebox{\textwidth}{!}{%
\begin{tabular}{|r|c|c|c|c|c|}
\hline
{\bf } & {\bf Total Accuracy} & {\bf Macro Precision} & {\bf Macro Recall} & {\bf Macro F-Measure} & {\bf Classification Time (ms)} \\ \hline
literature~\cite{machine-learning-ws-description-classification-2012} & 59.40 & 58.00 & 52.80 & 55.30 & unknown \\ \hline
cross-validation without context & 54.60 & 56.02 & 54.60 & 54.37 & 1478.00 \\ \hline
cross-validation with context & 59.00 & 59.65 & 59.00 & 58.62 & 3432.80 \\ \hline
\end{tabular}
}
\end{table}



The results demonstrate that \contextual{} information is improving the %
performance of the classification even though it is increasing the %
classification time due to increase in the file sizes because of the added %
\context{} information.

Finally, we use the best configuration (using \contextual{} information) in %
order to perform the final classification of all \wsdl{} files which class is %
unknown.
\xt{tab:WSDLSFinal} depicts the number of instances which was classified %
inside of each class.
These results are based on the performance of the current classification tool.
Currently, the results cannot be verified because the actual classes are not %
known. In order to validate, all the instances need be classified by human %
contribution by using approaches such as crowd-sourcing.

\begin{table}[htbp]
\centering
\caption{Final \wsdl{} classification results}
\label{tab:WSDLSFinal}
\begin{tabular}{|c|c|}
\hline
{\bf Category Name} & {\bf Number of Instances} \\ \hline
{\bf Weather} & 2782 \\ \hline
{\bf Social} & 2747 \\ \hline
{\bf Tourism} & 3581 \\ \hline
{\bf Financial} & 2376 \\ \hline
{\bf Entertainment} & 4110 \\ \hline
\end{tabular}
\end{table}


\subsection{WADL Classification Results}
\label{sec:WADLsClassificationResults}

\wadl{} descriptions are not popular %
through \serviceproviders{} and \rest{} \services{} are not widely described %
using \wadl{} in the \web{}.
As a result, we could not find %
as many instances for them as we could for \wsdl{} and \rest{} descriptions. 
The scale of the samples are not as much as the \wsdl{} and \rest{} samples %
and we discard the 10-fold cross-validation for these files. 
However, for the sake of completeness we perform a 2-fold cross-validation %
(swapping training and testing set and averaging the results) and we compare %
the results with the \context{} and without considering any \contextual{} %
information.

\xt{tab:WADLsCasesPrecision} depicts the highest accuracy achievable %
by testing algorithm permutations for each case of \wadl{} files using %
different clustering options.

\begin{table}[htbp]
\centering
\caption{Classification accuracy across data sets for \wadl{} files}
\label{tab:WADLsCasesPrecision}
\resizebox{\textwidth}{!}{%
\begin{tabular}{|c|c|c|c|c|c|c|}
\hline
 & {\bf \begin{tabular}[c]{@{}c@{}}TrainPlainCtx\\ TestPlainCtx\end{tabular}} & {\bf \begin{tabular}[c]{@{}c@{}}TrainPlain\\ TestPlain\end{tabular}} & {\bf \begin{tabular}[c]{@{}c@{}}TrainPlainCtx\\ TestPlain\end{tabular}} & {\bf \begin{tabular}[c]{@{}c@{}}TrainPlain\\ TestPlainCtx\end{tabular}} & {\bf \begin{tabular}[c]{@{}c@{}}TrainCtx\\ TestPlainCtx\end{tabular}} & {\bf \begin{tabular}[c]{@{}c@{}}TrainCtx\\ TestPlain\end{tabular}} \\ \hline
{\bf Mean} & 64.71 & 58.82 & 64.71 & 58.82 & 58.82 & 52.94 \\ \hline
{\bf No Clustering} & {\bf 76.47} & {\bf 64.71} & 58.82 & 52.94 & 58.82 & 47.06 \\ \hline
{\bf Median} & 58.82 & 58.82 & 58.82 & 58.82 & 52.94 & 41.18 \\ \hline
\end{tabular}
}
\end{table}

The highest accuracy without considering \context{} (Train on Plain-Test on %
Plain column) is achieved by using the \textit{No Clustering} option: %
64.71 percent.
The highest accuracy with considering \context{} (other columns) is achieved %
by using the \emph{No Clustering} option and using \emph{Plain + \Context{}} %
files in both training and testing: 76.47 percent.

\xt{tab:WADLsCrossValidated} depicts the result of two-fold cross-validation of the best cases.

\begin{table}[htbp]
\centering
\caption{\wadl s best cases cross-validated results}
\label{tab:WADLsCrossValidated}
\begin{tabular}{|c|c|c|}
\hline
 & {\bf \begin{tabular}[c]{@{}c@{}}TrainPlainCtx\\ TestPlainCtx\end{tabular}} & {\bf \begin{tabular}[c]{@{}c@{}}TrainPlain\\     TestPlain\end{tabular}} \\ \hline
{\bf No Clustering} & 63.235 & 62.91 \\ \hline
\end{tabular}
\end{table}


\subsection{REST Files Classification Results}
\label{sec:RESTsClassificationResults}

As discussed in \xs{sec:DataState}, because \html{} files contain too much %
noise, e.g., \emph{script} code, we define a new type of sample for \rest{} %
\html{} files and remove all the tags and unnecessary sections and only keep %
the raw text inside and store it in a separate text file.
We survey both sample types in order to find the best case for %
classification of \rest{} \service{} descriptions and use the sample type %
with the highest accuracy in the final classification.

\xt{tab:TEXTsCasesPrecision} depicts the highest accuracy achievable %
by testing algorithm permutations for each case of tags-filtered files describing %
\restful{} \services{} using different clustering options.

\begin{table*}[htbp]
\centering
\caption{Classification accuracy across data sets for \rest{} tags-filtered files}
\label{tab:TEXTsCasesPrecision}
\resizebox{\textwidth}{!}{%
\begin{tabular}{|c|c|c|c|c|c|c|}
\hline
 & {\bf \begin{tabular}[c]{@{}c@{}}TrainPlainCtx\\ TestPlainCtx\end{tabular}} & {\bf \begin{tabular}[c]{@{}c@{}}TrainPlain\\ TestPlain\end{tabular}} & {\bf \begin{tabular}[c]{@{}c@{}}TrainPlainCtx\\ TestPlain\end{tabular}} & {\bf \begin{tabular}[c]{@{}c@{}}TrainPlain\\ TestPlainCtx\end{tabular}} & {\bf \begin{tabular}[c]{@{}c@{}}TrainCtx\\ TestPlainCtx\end{tabular}} & {\bf \begin{tabular}[c]{@{}c@{}}TrainCtx\\ TestPlain\end{tabular}} \\ \hline
{\bf Mean} & 48 & 40 & 36 & 40 & 48 & 40 \\ \hline
{\bf No Clustering} & 44 & {\bf 48} & {\bf 52} & 44 & 48 & 40 \\ \hline
{\bf Median} & 40 & 44 & 48 & 40 & 44 & 40 \\ \hline
\end{tabular}
}
\end{table*}

The highest accuracy without considering \context{} (Train on Plain-Test on %
Plain column) is achieved by using the \textit{No Clustering} option: %
48 percent.
The highest accuracy with considering \context{} (other columns) is achieved %
by using the the \emph{No Clustering} option and using %
\emph{Plain + \Context{}} files in the training and \emph{Plain} files in the %
testing: 52 percent.
%
%
%



As the results depicted, the best case for the classification of \rest{} %
files is achieved by using the \emph{tags-filtered} sample type and training %
on \emph{Plain + \Context{}} files and testing on \emph{Plain} files and %
using the \emph{No Clustering} option.
As discussed in \xs{sec:TestingMethodology}, after finding the best case, we %
increase the sets and perform another exhaustive search in order to find the %
best algorithm combination. 

As the results depicted, the best result for the classification of \rest{} %
files is achieved by using the following configuration:

\begin{itemize}
	\item No Clustering option (discussed in \xs{sec:MARFOptions})
	\item \option{-silence-noise} for preparation, \option{-endp} for pre-processing, \option{-minmax} for feature extraction, \option{-cheb} for classification (discussed in \xs{sec:MARF})
\end{itemize}

As mentioned in \xs{sec:MARFOptions}, \option{-noise} removes noise
(high-frequency occurring material) by applying an FFT low-pass filter.
It is important to apply silence removal after the noise removal since
noise filtering may produce more silence gaps.
As a result, \option{-silence-noise} which combines the noise and silence 
removal helped selecting best low-frequency non-zero local minimums and 
maximum features in classification of these non-uniformly structured files.
Also, as mentioned in \xs{sec:MARFOptions} from a theoretical point of view, 
\option{-minmax} which picks a hundred features from the data, where 50 are 
minimums and 50 are maximums, works best with \option{-endp} in order to 
select the 100 local minimum and maximums.
In addition, Chebyshev distance classifier appears to work better with the higher-dimensionality of 100 features from 
\option{-endp} and \option{-minmax} selected local extremes due to their 
nature that provides enough discriminatory power for highly varied and 
overlapping \restful{} \services{}.

As discussed in \xs{sec:TestingMethodology} and similar to \wsdl{} files, %
we perform 10-fold cross-validation based on this configuration in order %
to give an insight on how the model will generalize to an independent dataset %
and to reduce variability.
\xt{tab:RESTsCrossValidated} depicts the cross-validated results including %
the evaluation measures which were defined in \xs{sec:TestingMethodology}.

\begin{table*}[htbp]
\centering
\caption{Effect of context on cross-validated \rest classification}
\label{tab:RESTsCrossValidated}
\resizebox{\textwidth}{!}{%
\begin{tabular}{|r|c|c|c|c|c|}
\hline
{\bf } & {\bf Total Accuracy} & {\bf Macro Precision} & {\bf Macro Recall} & {\bf Macro F-Measure} & {\bf Classification Time (ms)} \\ \hline
without context & 28.00 & 28.46 & 28.00 & 27.67 & 1801 \\ \hline
with context    & 29.60 & 29.67 & 29.60 & 29.12 & 3808 \\ \hline
\end{tabular}
}
\end{table*}

The performance is lower in comparison with \wsdl{} files due to the lack of common structure and high %
variability of these pages that describe \restful{} \services{} in %
different structures and terminologies and not in a structured format %
specific to describing \webservices{} like \wsdl{} files.
However, the performance is still significantly higher than the random baseline would have been.
Unlike \wsdl{} files, we could not find any related work for classification of %
\rest{} descriptions in the literature in order to compare with. 
As far as we know, this work is the initial step towards the classification %
of \rest{} descriptions.
We discuss in \xs{sec:ClassificationResultsEvaluation} how the performance %
of classification of \rest{} descriptions could be improved.  

In order to show the effect of \contextual{} information, as discussed in %
\xs{sec:TestingMethodology}, we perform another 10-fold cross-validation on %
the same configuration without considering any \contextual{} information and %
training on the \emph{Plain} files and testing on the \emph{Plain} files.
%
%

\xt{tab:RESTsCrossValidated} illustrates the effect of adding \contextual{} %
information to the the \rest{} \emph{tags-filtered} files on total accuracy, %
macro precision, macro recall, macro F-Measure, and classification time.


The results depict that \contextual{} information improves the %
performance of the classification even though it is increasing the %
classification time due to increase in the file sizes because of the %
\context{} which is added to them.
However, for \rest{} tags-filtered files it is not improving the accuracy %
as much as for \wsdl{} files due to the nature of these descriptions, %
which are not defined in a structured format specific to describing %
\webservices{}.
In other words, because they contain phrases which are more similar to the %
\contextual{} information phrases, \context{} is not adding much discriminant %
features to the \rest{} description files.

As a result, we use this configuration to perform the final classification of %
all \rest{} files.
We use the same training result which is the result of training on %
\emph{Plain + \Context{}} \rest{} \emph{tags-filtered} files and the %
\emph{Plain} \rest{} \emph{tags-filtered} files for the testing set to be %
classified.

Finally, we use the best configuration (using \contextual{} information) in %
order to perform the final classification of all \rest{} description files %
which class is unknown.
\xt{tab:RESTsFinal} depicts the number of instances, which were classified %
inside of each class.
Similar to the \wsdl{} files as discussed in \xs{sec:WSDLsClassificationResults}, %
these results are based on the performance of the current classification tool.
Currently, the full complete classification results cannot be verified because
the actual classes are not known.
In order to validate our complete data set, all the instances need be classified by human %
contribution by using approaches such as crowd-sourcing.

\begin{table}[htbp]
\centering
\caption{Final \rest{} classification results}
\label{tab:RESTsFinal}
\begin{tabular}{|c|c|}
\hline
{\bf Category Name} & {\bf Number of Instances} \\ \hline
{\bf Weather} & 6414 \\ \hline
{\bf Social} & 6323 \\ \hline
{\bf Tourism} & 5560 \\ \hline
{\bf Financial} & 6047 \\ \hline
{\bf Entertainment} & 6212 \\ \hline
\end{tabular}
\end{table}
%


\subsection{Evaluation}
\label{sec:ClassificationResultsEvaluation}

\wadl{} descriptions are not popular %
on the \web{} and \rest{} \services{} are not widely described %
using \wadl{} descriptions.
As a result, as discussed in \xs{sec:Classes}, because of the low number of %
\wadl{} files in the \repository{}, we were not able to find the same number %
of samples for the classes in comparison with other sample types.
The scale of the samples are not as much as the \wsdl{} and \rest{} samples %
and we discard the 10-fold cross-validation for these files. 
However, for the sake of completeness we performed a 2-fold cross-validation %
(swapping training and testing set and averaging the results) and we compared %
the results with the \context{} and without considering any \contextual{} %
information. Despite having to rely on limited data sets, our results show that %
the use contextual information does increase the effectiveness of \wadl{} %
classification. 

As the results depict, the accuracy is generally lower for \rest{} \html{} %
files in comparison with \wsdl{} files because they have more %
noise, e.g., \emph{JavaScript} and \emph{markup} code, and natural language segments. 
However, after filtering the tags and cleaning-up these files and storing the %
raw text inside in a separate text file, the accuracy increased in general %
and it helped the classification accuracy.
Although, the performance for them is still generally lower in contrast with %
\wsdl{} files due to the nature of these descriptions which are %
not defined in a structure format specific to describing \webservices{} and %
as a result, have high variability due to using different structures and %
terminologies embedded in the HTML documents.
However, the resulting performance is still higher than the random baseline after 10-fold %
cross-validation.
Unlike \wsdl{} files we could not find any related work for classification of %
\rest{} descriptions in the literature in order to compare with. 
As far as we know, this work is the initial step towards the classification %
of \rest{} descriptions.
The performance of the classification for these files can be improved using %
an approach to extract the most prominent features specific to these files %
before performing the classification. 
One way to achieve such goal is to extract all resource URIs using a regular expression %
extraction process.
However, because the URIs are also defined in different structures, formats, and shortcuts and have %
variability in the files, it requires significantly more experimentation to be done at the %
semantic-level processing.   

\emph{No Clustering} option, which was mentioned in \xs{sec:MARFOptions}, is %
found as the best clustering option in the best configurations of \marfcat{} %
for all of the three types of \service{} descriptions.
This option disables clustering the training and testing sets' individual %
class's feature vectors in \marf{}, i.e., it uses all of the feature vectors %
of the instances which we passed for a specific class in the training set %
and calculates their distance to all of the feature vectors of the %
instances which we passed for a specific class in the testing set instead of %
using only one feature vector (mean or median).
As a result, the space and the time complexity increases.
However, because our priority in finding the best configuration is the %
highest accuracy, we chose this option.

As the results depict, the algorithm combinations which are found as the %
best combinations in the best configurations of \marfcat{} vary throughout %
the different types of \service{} descriptions.
The reason is that this experiment is data-driven and the results is based on %
the input data. 
As a result, because the structure and nature of each of these types is %
different and also due to the manual choosing of training and testing sets %
for one type regardless of the other types, the aforementioned algorithm %
combinations vary.

The effect of adding \contextual{} information to the \wsdl{} files 
is illustrated in %
\xt{tab:WSDLsClassificationComparison}. 
The results depicts that \contextual{} information is improving the %
performance of the classification for both cases even though it is increasing %
the classification time due to increase in the file sizes because of the %
\context{} which is added to them.
The \context{} has less effect on the precision of \rest{} %
\emph{tags-filtered} files in comparison with \wsdl{} files due to the nature %
of these descriptions which are not defined in a structure format specific to %
describing \webservices{}.




\section{Concluding Summary}
\label{sec:ServiceClassificationSummary}
\label{sec:ResultsSummary}

In this article, we discussed our methodology of combining \machinelearning{} %
and Signal Processing techniques and employing \contextual{} information in %
order to automatically classify \webservice{} descriptions.
%
%
We defined 72 different cases based on the sample types, clustering options, %
and the \contextual{} information which we survey 864 combinations of %
algorithms and techniques in each. 



In \xs{sec:ClassificationResults} we measured and illustrated the results of %
\serviceclass{} including the resulting accuracies of 72 different cases %
based on the clustering options and adding \contextual{} information, %
cross-validated results for the best cases, the effect of adding %
\contextual{} information to the samples on the classification.
In addition, for \wsdl{} files we compared the same evaluation measures with %
the literature in order to give an insight on how close our classification %
process is to the literature.
Unlike \wsdl{} files we could not find any related work for classification of %
\rest{} descriptions in the literature in order to compare. 
As far as we know, this work is the initial step towards the classification %
of \rest{} descriptions.

We found and presented the best configuration of \marfcat{} (best algorithm %
combination + clustering option) which will result in the highest precision %
for each type of \service{} description.
In addition, we added the \contextual{} information to the classification and %
showed that it improves the performance of the classification and validated %
our hypothesis.

\bibliographystyle{alpha}
\bibliography{ws-classification-etc}

\end{document}